\documentstyle{article}

\input epsf

\begin{document}

\title{Interfaces between highly incompatible polymers of different stiffness:\\
Monte Carlo simulations and self-consistent field calculations} 

\author{M.\ M\"{u}ller${}^{1,2}$ and A. Werner${}^{1}$
\\
{\small ${}^1$ Institut f{\"u}r Physik, Johannes Gutenberg Universit{\"a}t} \\
{\small D-55099 Mainz, Germany}\\
{\small ${}^2$Department of Physics, Box 351560, University of Washington,} \\
{\small Seattle, Washington 98195-1560}}
\date{\today}
\maketitle

\begin{abstract}
We investigate interfacial properties between two highly incompatible polymers of different stiffness.
The extensive Monte Carlo simulations of the binary polymer melt yield detailed interfacial profiles
and the interfacial tension via an analysis of capillary fluctuations. We extract an effective Flory-Huggins
parameter from the simulations, which is used in self-consistent field calculations. These take due account
of the chain architecture via a partial enumeration of the single chain partition function, using chain 
conformations obtained by Monte Carlo simulations of the pure phases.
The agreement between the simulations and self-consistent field calculations is almost quantitative, 
however we find deviations from the predictions of the Gaussian chain model for high incompatibilities or
large stiffness. The interfacial width at very high incompatibilities is smaller than the prediction of 
the Gaussian chain model, and decreases upon increasing the statistical segment length of the semi-flexible 
component.
\end{abstract}

\section{ Introduction }
Melt blending of polymers has proven useful in designing new composite materials with improved
application properties. In many practical situations the constituents of the blend are
characterized by some degree of structural asymmetry. For example, a flexible component might contribute
to a higher resistance to fracture, while blending it with a stiffer polymer can increase the tensile 
strength of the material. Since the entropy of mixing in polymeric systems decreases with increasing degree of
polymerization, a small unfavorable mismatch in enthalpic interactions, entropic packing effects 
or the combination of both, generally 
leads to materials which are not homogeneous on mesoscopic scales, but rather fine dispersions 
of one component in another. Therefore properties of interfaces between unmixed phases are 
crucial in controlling the application properties of composites\cite{GENERAL} and have found
abiding experimental interest\cite{FRIEND1,FRIEND2,FRIEND3,FRIEND4}.

Recently, the bulk phase behavior and surface properties\cite{WU} of polyolefins\cite{BATES,OLEFINS} 
with varying microstructure
has attracted considerable experimental and theoretical interest. These mixtures are often modeled\cite{BATES,LIU,SCHWEIZER}
as blends of polymers with different bending rigidities, the less branched polymer corresponding
to the more flexible component. For pure hard core interactions, field theoretical calculations by Fredrickson, Liu and 
Bates\cite{LIU}, polymer reference interaction site model (P-RISM) computations by Singh and Schweizer\cite{SCHWEIZER}, lattice cluster theories 
by Freed and Dudowicz\cite{FREED} and Monte Carlo simulations\cite{M1} find a small positive contribution 
to the Flory-Huggins parameter $\chi$. Monte Carlo simulations which include a repulsion between unlike species reveal
an additional increase of the effective Flory-Huggins parameter with chain stiffness, because a back folding of 
chains becomes less probable with increasing stiffness and the number of intermolecular contacts increases\cite{M1} respectively. Qualitatively similar 
effects were found analytically in P-RISM\cite{SCHWEIZER} and lattice cluster\cite{FREED} theories.

In spite of their ubiquitous occurrence, interfacial properties in asymmetric blends have attracted comparably little 
interest. When entropic packing contributions to the Flory-Huggins parameter $\chi$ are small and composition
fluctuations are negligible, the self-consistent 
field theory is expected to yield an adequate quantitative description. Helfand and Sapse\cite{HS} extended the 
self-consistent field theory to Gaussian chains with different statistical segment lengths. 
In the limit of infinite long Gaussian chains and strong segregation, they obtained analytical expressions for 
the interfacial width $w$ and the interfacial tension $\sigma$. Both increase upon increasing the statistical segment
length of one component, leaving $\chi$ and the architecture of the other component unaltered. 

However, there are other models, that incorporate structural disparities on the monomer level.
Freed and coworkers model monomers as clusters of various shape on a lattice\cite{FREED2} and have explored corrections
to the energy of mixing and entropic contributions to the Flory-Huggins parameter.

Stiffness disparities have also been investigated using the worm-like chain model\cite{WORM},
which captures the crossover between rod-like behavior on small length scales and Gaussian
statistics on length scales much larger than the persistence length.
Morse and Fredrickson\cite{MORSE} extended the self-consistent field calculation to a symmetric 
blend of worm-like chains. For vanishing bending rigidity $\kappa$ they reproduced the Gaussian
chain result. In the limit of high bending rigidities and strong segregation ($\kappa\chi \gg 1$), however, they found 
that the width $w$ of the monomer density profile can be considerably smaller than for a Gaussian chain 
with the same long distance behavior. At large $\kappa \chi$, increasing the statistical segment length even leads to a 
decrease of the interfacial width in qualitative contrast to the Gaussian chain result. They also observed that the 
width of 
the bond orientation profile is of the order of the persistence length, which is much larger than $w$ in that limit.
Thus the interfacial width $w$ and the persistence length constitute two independent length scales
of the interfacial profiles. A reduction of the interfacial width in the case of small bending rigidities 
was obtained numerically by Schmid and M\"uller\cite{SCHMID1}. They noted that the local structure might become important
if its length scale is comparable to the interfacial width; a situation which occurs at rather large incompatibilities.

In the present study we extend our Monte Carlo studies\cite{M1} of structural asymmetric blends
to the investigation of interfacial properties between well segregated phases of
flexible and semi-flexible polymers. We consider rather small bending rigidities of the semi-flexible
component, so that the long distance behavior of both species is Gaussian. However, we chose the
incompatibility $\chi$ high enough, such that the interfacial width and the persistence length
are comparable for the higher bending rigidities. 
The Monte Carlo simulations highlight the architectural influences
and give a  detailed picture of interfaces between structural asymmetric polymers.
They yield density and orientation profiles for bonds and chains as a whole.
Extracting an effective Flory-Huggins parameter $\chi$ from the simulation data,
we compare our  Monte Carlo results to self-consistent field calculations 
which take due account of the chain architecture via a partial enumeration procedure\cite{SZLEIFER,M2,M2A}, 
and to Gaussian chain results. Therefore we can assess the importance of the level of coarse graining on the interfacial 
properties.

Our paper is organized as follows: 
In the next section we describe our polymer model, especially the dependence of single chain properties
on the stiffness. We comment on some computational aspects of the Monte 
Carlo simulations and describe the measurement of the interfacial tension. We also introduce the salient 
features of our self-consistent field calculations for arbitrary 
molecular architecture. In the following, we present our simulational results and compare them to the 
self-consistent field calculations. We close with a brief discussion of our findings and an outlook on future work.

\section{ Model and technical details }

\subsection{Bond fluctuation model and single chain properties}
In the framework of our coarse grained lattice model, a small number of chemical repeat units, say 3-5, is mapped 
onto a lattice monomer, such that the relevant features - chain connectivity and excluded volume interaction 
between monomeric units - are retained. We use the three dimensional bond fluctuation model (BFM)\cite{BFM}, 
which has found widely 
spread application in computer simulations, because it combines the computational efficiency of lattice 
models with a rather faithful approximation of continuous space properties. 
Each effective monomer blocks 
a cube of 8 neighboring sites from further occupancy on a simple cubic lattice. 
Due to the extended monomer size, the
model captures some nontrivial packing effects. 
We consider a blend of $n_A$ flexible polymers of length $N_A$ and $n_B$ semi-flexible B-polymers comprising $N_B$
monomers in a volume $V$.
At a total monomer density
$\Phi_0 = (N_An_A+N_Bn_B)/V= 0.5/8$, the model reproduces many properties of a dense polymeric melt. 
We use chain lengths $N=N_A=N_B=32$
and $64$, which correspond to a degree of polymerization of the order 120 and 240 in more chemically realistic polymer models.  
Monomers are connected via 
one of 108 bond vectors with lengths $2,\sqrt{5},\sqrt{6},3$ or $\sqrt{10}$, where here, and henceforth, all lengths are 
measured in units of the lattice spacing. The large number of bond vectors permits 87 different bond angles. 

The persistence 
length of the semi-flexible B-polymers is tuned by imposing an intermolecular potential, which favors straight bond 
angles. We use a particular simple choice\cite{M1}:
$E(\theta) = f k_BT \cos(\theta)$
where $\theta$ denotes the complementary angle to two successive bonds.
Previous Monte Carlo simulations\cite{M1}
of the bulk thermodynamics for $N=32$ and $f=1.0$ revealed a purely entropic Flory-Huggins parameter 
$\Delta \chi = 0.0018(2)$ for the athermal blend. This small value is in good quantitative agreement with 
theories\cite{LIU,SCHWEIZER}. These packing effects result in a slight increase of the osmotic pressure
with the bending energy, which gives rise to a monomer density difference of about $1\%$ between the coexisting phases.

Since $\Delta \chi \ll 2/N = 0.0625$ for this combination of chain length and stiffness disparity, we introduce
an additional enthalpic repulsion to induce phase separation. For simplicity, these thermal interactions are modeled
as a square well potential comprising all 54 neighbor sites up to a distance $\sqrt{6}$. The contact of monomers 
of the same species lowers the energy by $\epsilon k_BT$, whereas the contact of different monomers increases the energy 
by the same amount. 
A finite size scaling analysis yields accurate estimates for the critical point
of the binary blends ($N=32$):
$\epsilon_c=0.01442(6), \phi_{Ac}=0.5$ and $\epsilon_c=0.0127(1), \phi_{Ac}=0.516(10)$ for 
$f=0$ and $1$\cite{M1}, respectively.
In the present study we chose $\epsilon=0.05$ which corresponds roughly to $\chi \approx 0.27$. 
This value is much higher than typical values for polyolefin blends\cite{BATES}. 
Our results correspond to rather strongly immiscible blends (e.g.\ interfaces between polystyrene (PS) and 
polyvinylpropylene PVP \cite{FRIEND4}).

\begin{figure}[htbp]
    \begin{minipage}[t]{160mm}%
       \setlength{\epsfxsize}{13cm}
       \mbox{\epsffile{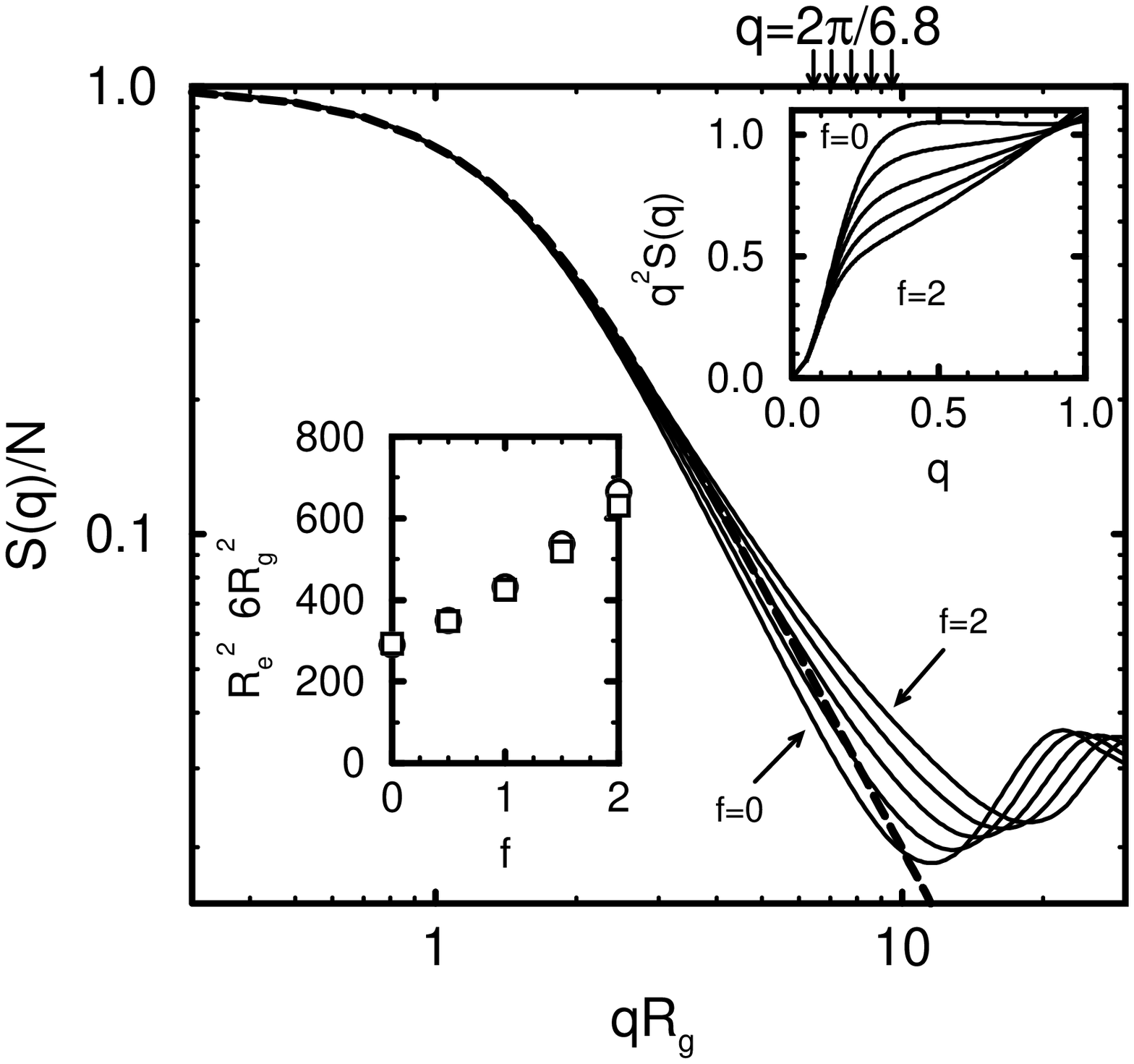}}
    \end{minipage}%
    \hfill%
    \begin{minipage}[b]{160mm}%
       \caption{
       Single chain structure function $S(q)$ for bending energies $f=0.0, 0.5, 1.0, 1.5,$ and $2.0$ (solid lines).
       The dashed line represents the Debye function. The arrows, on top of the figure, characterize the
       interfacial length scale ($2w_{\rm SCF} = 6.8$).
\newline Left inset:
       Mean squared end-to-end distance (circles) and radius of gyration (squares) as a function of the bending energy $f$.
       Upon increasing $f$, the chain extension grows about a factor 1.5.
       Note that the ratio $R_e^2/R_g^2$ remains close to its Gaussian value 6. 
\newline Right inset:
       Kratky plot of the single chain structure factor $S(q)$ for $f=0.0, 0.5, 1.0, 1.5,$ and $2.0$ from top to
       bottom. A plateau, characteristic of Gaussian statistics, is only found for $f=0$ (flexible chains), whereas
       there are pronounced deviations for semi-flexible chains.
               }
       \label{fig:conf}
    \end{minipage}%
\end{figure}

The conformational data for $N=32$ and $\epsilon=0.05$ as a function of the bending energy $f$ are presented in Fig.\ \ref{fig:conf} and Table \ref{tab:konf}.
The inset shows the growth of the chain extension upon increasing $f$. The ratio between the square end-to-end distance $R^2$
and the square radius of gyration $R_g^2$ remains very close to the Gaussian value $6$ (within $5\%$ even for $f=2$). Also the small 
wave vector regime of the single chain structure function 
\begin{equation}
S(q)=\frac{1}{N} \left \langle \left| \sum_{i=1}^{N} \exp(i\vec{q}\vec{r}_i)\right|^2 \right \rangle
\end{equation}
is well describable by a Debye function $S_D(q)/N = 2[\exp(-q^2R_g^2)-1+q^2R_g^2]/(q^2R_g^2)^2$\cite{FLORY} for $q<0.3$. Thus the {\em long} 
range behavior of
our chains is characterized by Gaussian statistics for all values of the bending energy $f$ studied and we define the statistical
segment length $b$ according $b^2 = R^2/(N-1)$. Note that the statistical segment length grows from 3.06 for $f=0$ to $4.63$
for $f=2$. This asymmetry in the statistical segment length is of similar magnitude as in polyolefin blends\cite{BATES}.

However, for length scales of the order of the statistical segment length, we find deviations from
the Gaussian behavior.  The plateau $q^2S(q)=12/b^2$ for large $q$ in the Kratky 
Porod plot is only observed for flexible chains ($f=0$) and yields a slightly higher estimate for the
statistical segment length $b=3.4$. For the semi-flexible chains the slope of $q^2S(q)$ in the range 
$0.3<q<1$ increases upon increasing the bending energy. Defining an effective bending rigidity of an 
equivalent worm-like chain
$\kappa = R^2/2\langle b^2\rangle (N-1)$ ($\langle b^2\rangle$: mean squared bond length), 
$\kappa$ grows from 0.68 to 1.57 ($\kappa\chi = 0.18 \cdots 0.47$) upon increasing the bending energy $f$.
For wave vectors $q R_g\approx 2\pi R_g/2 w$ (denoted by the arrows in the Fig.\ \ref{fig:conf}), 
where $w=3.4$ corresponds roughly to the width of the monomer density profile in the self-consistent 
field (SCF) calculations, we find deviations from the Gaussian behavior for higher bending energies and anticipate corrections 
to the predictions of the Gaussian model.

\subsection{Local fluid structure and effective Flory-Huggins parameter $\chi$}
In order to compare our simulational results to self-consistent field (SCF) calculations, which cannot account for the local fluid
structure of our model, we have to identify an effective Flory-Huggins parameter $\chi$. For the bulk behavior in 
the one phase region this has been discussed in ref.\ \cite{M1,M0}:
We define a dimensionless monomer density $\phi_{A(B)}$ as the ratio between the local number density of A(B)-monomers
and the total monomer density $\Phi_0$. Then, the density of intermolecular contacts $n_{AB}$ takes the form:
\begin{equation}
\frac{2n_{AA}}{\Phi_0\phi_A^2} = \Phi_0  \int_{r \leq \sqrt{6}}d^3r\; g_{AA}(r) \equiv z_{AA}
\qquad \mbox{and} \qquad
\frac{ n_{AB}}{\Phi_0\phi_A\phi_B} = \Phi_0 \int_{r \leq \sqrt{6}}d^3r\; g_{AB}(r) \equiv z_{AB}
\end{equation}
where $g_{IJ}$ denotes the $IJ$ interchain correlation function, which is normalized such that $g_{IJ}(r \to \infty)=1$.
The integration is extended over the spatial extension of the square well potential and $z_{IJ}$ corresponds to the
effective coordination number of the Flory-Huggins lattice. If the coupling between chain conformations and 
effective monomer repulsion is negligible, only the {\em inter}molecular energy drives the phase separation.
In this case (as we shall see in the next subsection), the $\chi$ parameter takes the form:
$\chi=\epsilon(z_{AA}+2z_{AB}+z_{BB})/2k_BT$, where $z_{IJ}$ denote the coordination numbers obtained from the
{\em inter}molecular pair-correlation functions. At the critical temperatures the coordination numbers
have been measured in the simulations at composition $\phi_A=\phi_B=1/2$\cite{M1}.
From its very definition the Flory-Huggins parameter $\chi$ accurately describes the
intermolecular interaction energy, and it agrees nicely with values obtained from the semi-grandcanonical
equation of state and the estimate from the long wavelength behavior of the collective structure factor.
It also yields estimates of the critical temperature, which agree with the Monte Carlo results up to $1/\sqrt{N}$ 
corrections due to composition fluctuations\cite{M0}. 

\begin{figure}[htbp]
    \begin{minipage}[t]{160mm}%
       \setlength{\epsfxsize}{13cm}
       \mbox{\epsffile{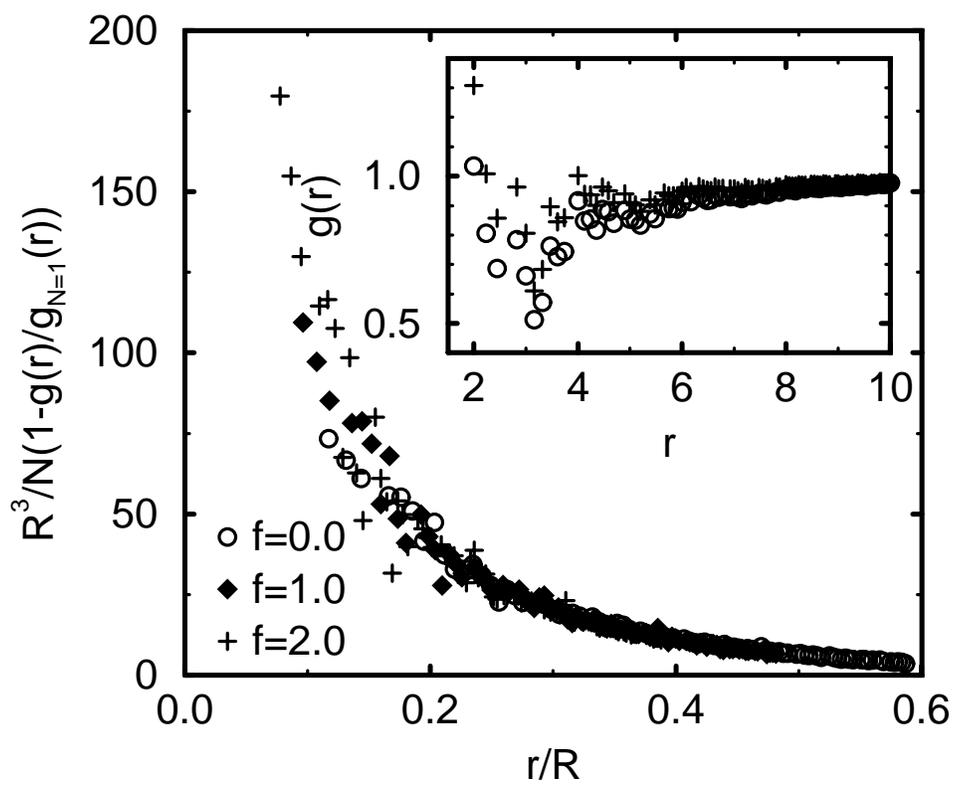}}
    \end{minipage}%
    \hfill%
    \begin{minipage}[b]{160mm}%
       \caption{Scaling plot for the intermolecular pair-correlation function for bending energies $f=0,\;1$ and $2$.
		The inset shows the pair-correlation function for the lowest and the highest bending energy
		studied.
               }
       \label{fig:ginter}
    \end{minipage}%
\end{figure}

In the pure system, the intermolecular coordination number of the flexible component is lower than the 
corresponding value for the semi-flexible chains\cite{M1}. The number of intramolecular contacts\cite{C2} is higher for the flexible chains.
Therefore, the $\chi$-parameter grows upon increasing
$f$\cite{M1}. Due to the larger chain extension for the semi-flexible component, the correlation hole has 
a larger spatial extent, but is more shallow. The intermolecular pair correlation function is presented in
the inset of Fig.\ \ref{fig:ginter}. Due to the extended monomer size $g(r)$ vanishes for distances $r<2$. At short
distances, the presence of single site vacancies introduces local packing effects, which gives rise to 
several neighbor shells in the fluid. The extended structure of the polymer manifests itself in a reduction
of contacts with {\em other} chains on the length scale of the end-to-end distance. On short distances, the intermolecular
pair-correlation function for the stiffer chains is larger than for the flexible ones.
For flexible chains
it is possible to separate the monomeric packing effect from the polymeric correlation hole by dividing 
$g(r)$ by its monomeric equivalent\cite{M0}, which exhibits only packing effects. The ratio $g(r)/g_{N=1}(r)$ presents
the conditional probability of finding a monomer of a different chain at a distance $r$, if there would be one
in the monomer system. This ratio, presented in Fig.\ \ref{fig:ginter}, is a rather smooth function, indicating, that 
the chain connectivity hardly affects the monomeric packing. If the correlation hole would be characterized
by a single length scale, i.e.\ the end-to-end distance $R$ in the Gaussian chain model, one expects a
scaling behavior of the form:
\begin{equation}
1-\frac{g(r)}{g_{N=1}(r)} = \frac{N}{R^3} f\left(\frac{r}{R}\right)
\end{equation}
Such a scaling plot is shown in Fig.\ \ref{fig:ginter}. The data collapse well for the different bending rigidities
at large distances, whereas there are deviations for small distances. This is a further indication,
that the chain structure is characterized by two independent length scales, the end-to-end distance and the
persistence length.

In the well segregated regime (far below the critical temperature), it is very difficult to measure the $AB$ 
intermolecular correlation function in the bulk. Therefore, unlike ref.\ \cite{M1}, we make an {\em additional}
ad-hoc assumption: $z_{AB}=(z_{AA}+z_{BB})/2$. For symmetric blends near the critical point P-RISM calculations\cite{YETH}
predict that deviations from this behavior die out with growing chain length like $1/\sqrt{N}$.
However, the validity of this random-packing like assumption for highly incompatible structural 
asymmetric blends is not obvious. 

We explore the interfacial structure by simulating a system in a $L\times L\times 2L$ geometry with $L=64$ and periodic boundary 
conditions in the canonical ensemble. The system
contains two interfaces parallel to the $xy$ plane. The chain conformations are generated via local monomer displacements and 
slithering snake moves, which are applied at a ratio 1:3 (except for $f=0$, where only local monomer displacements were employed).
The systems were equilibrated over 125,000 attempted local moves per monomer (AMM) and  375,000 slithering  snake
tries per chain (SS).  Every 12,500 AMM and 37,500 SS movements a configuration was stored for detailed analysis,
at least 898 configuration were generated. We use a trivial parallelization strategy on a CRAY T3E, running typically
8 or 32 configurations in parallel.

Profiles across the interface are measured according to the following procedures:
``Apparent'' profiles are obtained by locating the instantaneous position of the interface across the whole
lateral system extension in each snapshot and averaging over profiles with respect to the instantaneous, but 
laterally averaged midpoint. These profiles exhibit a system size dependent broadening due to capillary fluctuations, 
which is not accounted for in the SCF calculations. 
To avoid this broadening, we 
define ``reduced'' profiles by laterally dividing the system into subsystems of size $B \times B$. We choose $B=16$.
One could reduce the effect of capillary fluctuations further by chosing a smaller block size $B$, however, one should 
take care not to cut off ``intrinsic'' fluctuations\cite{SEM_CAP}. Since on the scale $B$  fluctuations are still 
reasonably described by a Helfrich Hamiltonian\cite{HELFRICH}(see below), our block size $B$ is larger than the 
length scale of ``intrinsic'' fluctuations. This is consistent with Semenov's\cite{SEM_CAP} estimate for the corresponding 
length scale $L_{\rm cutoff}=\pi w \approx 10<B$.
Thus this averaging procedure reduces the influence of capillary fluctuations, but does not eliminate it completely\cite{AW}.

\begin{figure}[htbp]
    \begin{minipage}[t]{160mm}%
       \setlength{\epsfxsize}{13cm}
       \mbox{\epsffile{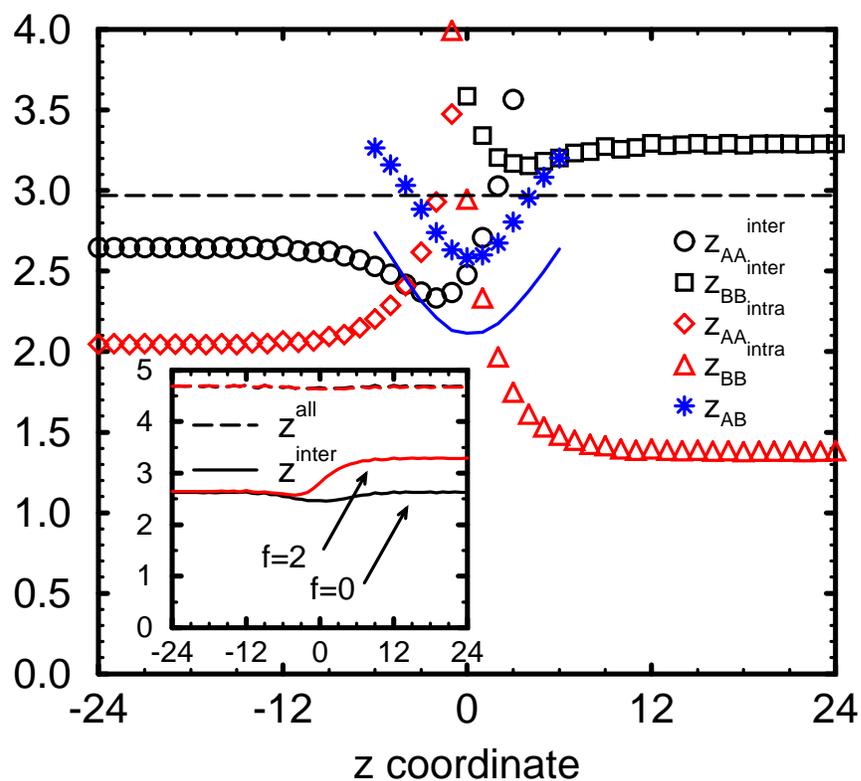}}
    \end{minipage}%
    \hfill%
    \begin{minipage}[b]{160mm}%
       \caption{Reduced profiles of the coordination numbers of inter- and intrachain contacts for the bending energy $f=2$
		The flexible component is on the left side of the interface. The line presents the apparent profile for
		the AB intermolecular contacts. The dashed line is the random mixing like approximation for the
		AB intermolecular contacts. The inset shows the total number of contacts and number of interchain
		contacts.
               }
       \label{fig:zcoord}
    \end{minipage}%
\end{figure}

The presence of an interface gives rise to a spatial dependence
of the local monomer densities and chain conformations, which in turn is reflected in the intermolecular pair 
correlation functions. In Fig.\ \ref{fig:zcoord} we present the reduced profiles of the intermolecular and 
intramolecular coordination numbers as a function of the distance from the center of the interface for the 
bending energies $f=0$ and $2$. The individual coordination numbers exhibit a considerable spatial dependence; 
this is however partially due to the spatial range of interactions and the remaining capillary fluctuations\cite{C1}.
To illustrate the effect we plot the apparent and reduced profile of the AB intermolecular coordination number.
The value at the center of the interface increases upon reducing $B$; 
the intrinsic value can not be estimated from these data with high precision. However, the average value of the
``reduced'' profile
is close to $z_{AB}=(z_{AA}+z_{BB})/2$, the value used in the SCF calculations.

The total number of intermolecular contacts $z^{inter}=(n_{AA}^{inter}+n_{BB}^{inter}+n_{AB}^{inter})/
\Phi_0(\phi_A+\phi_B)^2$ is much less sensitive to the intrinsic (local) profiles and
shows a gradual transition between the corresponding bulk values, with a reduction at the center of
the interface\cite{M4} of about $8\%$. This spatial dependence of the effective Flory Huggins parameter
is neglected.
Interestingly, the sum of all contacts $z^{all}$ (both intermolecular and intramolecular)
is largely independent of the stiffness or the distance from the interface, i.e.\ the bending energy or the
unfavorable interactions at the interface causes the chains to rearrange (e.g.\ exchange unfavorable interchain
contacts by energetic favorable intrachain contacts) but hardly affect the structure of the underlying monomer fluid.

Therefore, the local fluid structure is dominated by the packing constraints and the excluded volume interactions.
The chain connectivity, bending energies, and the thermal interactions are of minor importance for the monomer fluid.
The chain conformations are strongly influenced by the bending energies but depend only slightly on the thermal
interactions. The Flory-Huggins parameter is determined by the thermal interactions and also depends on the 
bending energies via the correlation hole effect. The disparity in the packing behavior of the flexible and 
the stiff polymers is of minor importance for $\chi$ for the chain lengths studied.

\subsection{Measuring the surface tension via the capillary fluctuation spectrum}

\begin{figure}[htbp]
    \begin{minipage}[t]{160mm}%
       \setlength{\epsfxsize}{13cm}
       \mbox{\epsffile{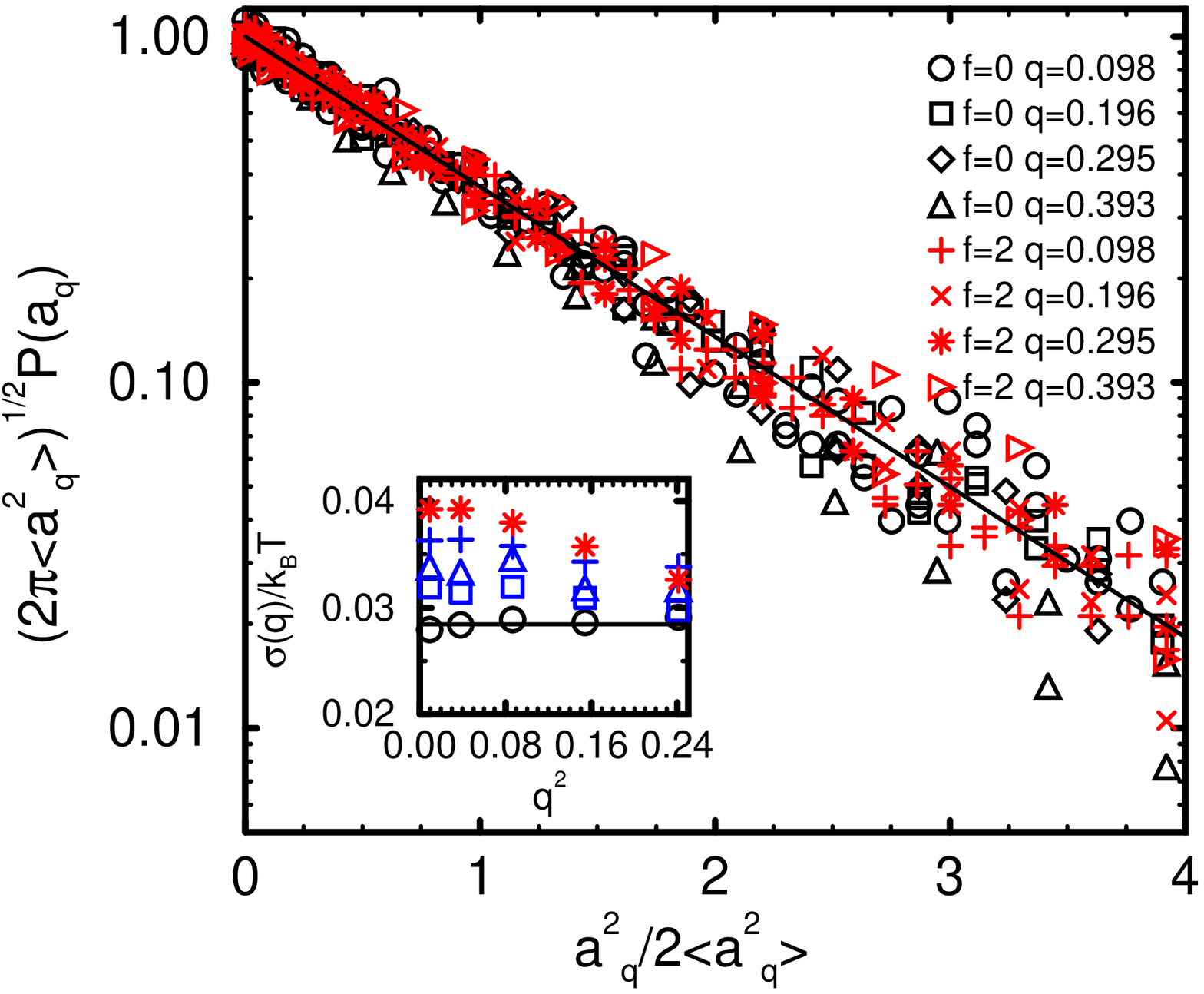}}
    \end{minipage}%
    \hfill%
    \begin{minipage}[b]{160mm}%
       \caption{
       Probability distribution of Fourier components of the local interfacial profile for the 4 smallest
       wave-vectors $q$ and bending energies $f=0$ and $2$, using the $1/q^2$ dependence of the variance. 
       The solid line represents the expected Gaussian distribution.
\newline inset:
       Inverse width of the distribution	$\sigma(q) = \frac{2}{L^2q^2\langle a_q^2\rangle}$ for all bending
       energies (circles $f=0.0$, squares $f=0.5$, triangles $f=1.0$, crosses $f=1.5$, and stars $f=2.0$) and 
       the smallest $q$ values. The 3 smallest wave-vectors were used to determine the interfacial tension $\sigma$. 
       The Helfrich Hamiltonian describes the data down to smaller $q$ values for the symmetric blend than for $f=2$.
       The solid line denotes the independent estimate for the symmetric blend ($f=0$) from simulations in the 
       semi-grandcanonical ensemble.
               }
       \label{fig:sigma_all}
    \end{minipage}%
\end{figure}

Due to the stiffness disparity between the species, straightforward application of semi-grandcanonical identity 
changes between 
different polymer types are rather inefficient (note that the efficiency drops by about 3 orders of magnitude\cite{M1} 
upon increasing $f$ from 0 to 1 for $N=32$) and therefore limited to small chain length and stiffness. Measurement of the 
interfacial
tension via the reweighting of the composition distribution, which has been successfully applied to structural symmetric
blends ($f=0$), is therefore difficult. In principle, the interfacial tension can be determined via the anisotropy of the 
pressure tensor. This method has been successfully applied in off-lattice simulations\cite{PRESSURE}, but the 
generalization to lattice models is difficult\cite{CIFRA}. However, the spectrum of capillary fluctuations  offers an 
alternative\cite{M3} for measuring the interfacial tension; a method which does not rely on identity switches.
Let $u(x,y)$ be the local interfacial position. Then the free energy 
cost for deviations from a flat planar interface is given by the Helfrich expression\cite{HELFRICH}:
\begin{equation}
{\cal H} = \int dxdy\; \frac{\sigma}{2} (\nabla u)^2+ \cdots
\end{equation}
where higher order gradient terms are neglected. In our simulation, we define local $x$- and $y$-averaged interface positions by 
minimizing the quantity
\begin{equation}
\left| \sum_{z=u(y)-6}^{u(y)+6}\sum_{x=0}^{x=L-1} {\Large (}\phi_A(x,y,z)-\phi_B(x,y,z) {\Large )}\right|
\end{equation}
for the $x$-averaged position $u(y)$ and a similar expression for the $y$-averaged one. This averaged interfacial position 
is Fourier decomposed according to:
$
u(y)=\frac{a_0}{2} + \sum_{l=0}^{L/2} a(q_l)\cos(q_ly) + b(q_l)\sin(q_ly)
$
with $q_l=2\pi l/L$. The Helfrich Hamiltonian predicts that the Fourier components $a(q_l)$ and $b(q_l)$ are Gaussian 
distributed with a width
\begin{equation}
\frac{2}{L^2\left\langle a^2(q)\right\rangle} = \frac{2}{L^2\left\langle b^2(q)\right\rangle} = \frac{\sigma}{k_BT} q^2
\end{equation}
In Fig. \ref{fig:sigma_all} we present the distribution of the Fourier components for two different bending energies $f=0,2$ and the 
4 smallest wave vectors $q$. This long wavelength part of the fluctuation spectrum is well described by the 
quadratic Helfrich expression. The straight line marks the expected Gaussian distribution for the Fourier amplitudes,
to which the simulation data comply. The inverse width of the distribution determines the interfacial tension.
The extracted value for the symmetric blend agrees with the independent measurement obtained via the reweighting
scheme\cite{M4}. (The latter scheme measures the interfacial free energy via the ratio of the probability for finding
the system in a homogeneous bulk state or a configuration comprising two interfaces. )
To estimate the errors of measuring the interfacial tension via the capillary fluctuation spectrum, it would be desirable to increase the lateral system size.
However, the error in extrapolating the simulation data to $q \to 0$ is smaller than $7\%$.
Thus the analysis of the capillary fluctuation spectrum is an efficient alternative for measuring interfacial
tensions in structurally asymmetric systems; the results are compared to the predictions of the SCF
calculations in Sec.\ IIIa.

\subsection{Self-consistent field calculations}
The mean field  approach is similar to Helfand\cite{HT,HS}, Noolandi\cite{NOOLANDI}, and Shull\cite{SHULL}, except for the treatment of
the chain architecture\cite{M2}. The partition function of a binary polymer blend has the general form\cite{H75}:
\begin{equation}
{\cal Z} \sim \frac{1}{n_A!n_B!} 
	      \int \Pi_{\alpha=1}^{n_A} {\cal D}[r_\alpha] {\cal P}_A[r_\alpha]
	           \Pi_{\beta=1}^{n_B} {\cal D}[r_\beta] {\cal P}_B[r_\beta]
	      \exp \left( -\frac{\Phi_0}{k_BT} \int d^3r\; {\cal E}(\hat{\phi}_A,\hat{\phi}_B)\right)
\end{equation}
where the functional integrals ${\cal D}[r]$ sum over all polymer conformations and ${\cal P}[r]$ denotes
the probability distribution characterizing the noninteracting, single chain conformations. ${\cal E}$ represents a 
segmental interaction free energy, and the dimensionless monomer density takes the form\cite{H75}:
\begin{equation}
\hat{\phi}_A(r) = \frac{1}{\Phi_0} \sum_{\alpha=1}^{n_A}\sum_{i_A=1}^{N_A} \delta(r-r_{\alpha,i_A})
\end{equation}
where the sum runs over all monomers in the A-polymer $\alpha$. A similar expression holds for $\hat{\phi}_B(r)$.

The segment free energy ${\cal E}$ comprises two contributions:
a free volume part arising from hard core interactions and an energetic term from the thermal interactions. 
Since the melt is nearly incompressible, we approximate the free volume
part by a simple quadratic expression introduced by Helfand\cite{HT}, which  reproduces the relative reduction of the total
monomer density by about $4\%$\cite{SCHMID1}. However the difference of the bulk densities of the coexisting phases has a different sign
in the simulations and than in the SCF calculations. In the simulations the higher osmotic pressure of the semi-flexible component results 
in a slightly lower bulk density of the 
semi-flexible component in the simulations, an effect neglected in the SCF calculations.
Moreover, in the SCF calculations the more negative intermolecular energy density (see below) of the B component
results in a slightly higher density of semi-flexible polymers. The total density differences between the
coexisting phases is however only about $1\%$.
The pairwise intermolecular interactions $V_{IJ}(r)$ ($I,J$=A,B) are treated as point interactions 
of strength $\epsilon z_{IJ} \delta(r)/\Phi_0$. $z_{IJ}$ parameterizes the local fluid structure of the underlying 
microscopic model, as discussed above. The coupling between individual chain conformations and the coordination numbers, 
which results in the spatial dependence of the $\chi$-parameter observed in the simulations, is neglected.
Furthermore, we ignore purely entropic contributions (which have been determined to be small by Monte Carlo simulations) 
and do not
include orientation dependent segmental interactions, which will eventually lead to a nematic phase at much higher bending
energies $f$. Thus we take the interactions to be
\begin{equation}
\frac{{\cal E}(\phi_A,\phi_B)}{k_BT} =   \frac{\zeta}{2} \left( \phi_A + \phi_B - 1\right)^2 
		    - \frac{\epsilon z_{AA}}{2} \phi_A^2
		    - \frac{\epsilon z_{BB}}{2} \phi_B^2
		    + \epsilon z_{AB} \phi_A \phi_B
\end{equation}
The inverse compressibility $\zeta$ has been measured in simulations of the athermal model; $\zeta=4.1$\cite{WM1}.
A Hubbard-Stratonovich transformation rewrites the many chain problem in terms of independent chains in external, 
fluctuating fields $W_A$ and $W_B$.
\begin{equation}
{\cal Z} \sim \int {\cal D}[W_A,W_B,\Phi_A,\Phi_B] \exp \left(-{\cal F}[W_A,W_B,\Phi_A,\Phi_B]/k_BT\right)
\end{equation}
where the free energy functional is defined by
\begin{eqnarray}
\frac{{\cal F}[W_A,W_B,\Phi_A,\Phi_B]}{\Phi_0 k_BT V} &=& 
					    \frac{\bar{\phi}_A}{N_A} \ln \bar{\phi}_A
					 +  \frac{\bar{\phi}_B}{N_B} \ln \bar{\phi}_B
					 +  \frac{1}{V} \int d^3r \; {\cal E}(\Phi_A,\Phi_B)  \nonumber \\
		&&   	                 -  \frac{1}{V} \int d^3r \left\{  W_A\Phi_A + W_B\Phi_B \right\}
					 -  \frac{\bar{\phi}_A}{N_A} \ln q_A[W_A]
					 -  \frac{\bar{\phi}_B}{N_B} \ln q_B[W_B]
\end{eqnarray}
$\bar{\phi}_A = \frac{n_A N_A}{\Phi_0 V} = 1 -\bar{\phi}_B$ denotes the average A-monomer density and $q_A[W_A]$ the
single chain partition function in the external field $W_A$
\begin{equation}
q_A[W_A] = \frac{1}{V} \int {\cal D}_1[r] {\cal P}_A[r] \exp \left(- \Phi_0 \int d^3r\; \hat{\phi}_A W_A \right)
\end{equation}
respectively. The leading contributions to the partition function stem from those values $\phi_A,\phi_B,w_A,w_B$ of the
collective variables which extremize the free energy functional, and the mean field approximation amounts to retaining 
only these contributions. The values are determined by:
\begin{eqnarray}
\frac{\delta {\cal F}}{\delta \phi_A} = 0 &\Rightarrow &
w_a = \frac{\delta}{\delta \phi_A} \int d^3r\; {\cal E}(\phi_A,\phi_B) =
\zeta (\phi_A + \phi_B -1) - \epsilon z_{AA} \phi_A + \epsilon z_{AB} \phi_B \\
\frac{\delta {\cal F}}{\delta w_A} = 0    &\Rightarrow &
\phi_A = \frac{\bar{\phi}_A V}{N_A q_A} \frac{\delta q_A}{\delta w_A} \label{eq:d}
\end{eqnarray}
and similar expressions for $w_B$ and $\phi_B$. 
The saddle point integration approximates the original problem of 
mutually interacting chains by one of a single chain in an external field, which is determined, in turn, by the monomer density.
Composition fluctuations are ignored,
but the coupling between chain conformations (e.g.\ orientations) and the monomer density is retained.
The free energy of a homogeneous system takes the Flory-Huggins form:
\begin{equation}
\frac{{\cal F}}{\Phi_0k_BTV} = \frac{\bar{\phi}_A}{N_A}\ln\left(\bar{\phi}_A\right)
			      +\frac{1-\bar{\phi}_A}{N_B}\ln \left(1-\bar{\phi}_A\right)
			      -\frac{1}{2}\epsilon\left\{ \left( z_{AA}+2z_{AB}+z_{BB}\right)\bar{\phi}_A^2
							  -2\left(z_{AB}+z_{BB}\right)\bar{\phi}_A +z_{BB}
						  \right\}
\end{equation}
where we identify the Flory-Huggins parameter $\chi=( z_{AA}+2z_{AB}+z_{BB} )\epsilon/2$.
In the strongly segregated regime, the free energy of a system containing one interface is given by:
${\cal F}=-\epsilon(\bar{\phi}_Az_{AA}+(1-\bar{\phi}_A)z_{BB})\Phi_0k_BTV/2+\sigma k_BTL^2$.
The definition of the interfacial tension as the difference of the free energy of a system containing an interface
and the homogeneous bulk system corresponds literally to the measurement of the interfacial tension via the reweighting 
scheme\cite{M4} in the Monte Carlo simulations. As shown in Sec. IIc for the symmetric blend, these values agree with 
the measurement of the interfacial tension via the capillary fluctuation spectrum, so that we can compare the 
results of the SCF calculations and the values extracted from the capillary wave spectrum quantitatively.

For the special cases of Gaussian chains\cite{HT,EDWARDS} and worm-like polymers\cite{WORM,MORSE} one can treat the 
single chain problem in an arbitrary external field in limiting cases (e.g\ $N \to \infty$) analytically.
For general parameters, however, one has to resort to numerical procedures even for these simple models.
The BFM chains used in the simulations are characterized by structure on different length scales.
The conformations are rod-like for length smaller than the persistence length, which
depends on the bending energy $f$. On intermediate length scales, they obey self-avoiding walk statistics, while
on the largest scale, the excluded volume interactions are screened in the melt, and the chains exhibit Gaussian
statistics. Since we want to explore dependence on the explicit chain structure, we evaluate the single chain
partition function via a partial enumeration scheme, introduced by Szleifer and coworkers\cite{SZLEIFER}. The method 
is conceptually straightforward and applicable to {\em arbitrary} architecture\cite{M2,M2A}. It can use experimental or 
simulational data as input.
Note that no adjustable parameters are involved in the chain structure (such as the statistical segment length in 
the Gaussian model or the bond length and the bending rigidity in the worm-like polymer model) and the chain structure 
is correctly represented on {\em all} length scales.
Using Monte Carlo simulations of the pure melt, we generated 40,960 independent polymer 
conformations for each bending energy. Rotating 
and translating those original conformations, we obtain a sample of 7,864,320 
polymer conformations for chain length $N=32$. 
(Note only the $z$ coordinates of the chains are employed for a  flat
interface parallel to the $xy$ plane.)
For $N=64$ we use twice as many conformations.
Within this framework, the A-monomer density (c.f.\ eq.\ \ref{eq:d}) is simply the statistical average of independent
A-polymers in the external field $w_A$:
\begin{equation}
\phi_A = \bar{\phi}_A \frac{   \sum_{\alpha=1}^{C} \frac{1}{N_A} \sum_{i=1}^{N_A} V \delta(r-r_{\alpha,i}) 
			       \exp \left( -\sum_{i=1}^{N_A} w_A(r_{\alpha,i})  \right)                       }
                           {   \sum_{\alpha=1}^{C}
			       \exp \left( -\sum_{i=1}^{N_A} w_A(r_{\alpha,i})  \right)                       }
\end{equation}
Other single chain quantities are given by corresponding averages over independent chains in the fields $w_A$ and $w_B$. 

The set of nonlinear equations is expanded in a Fourier series\cite{MATSEN} and solved by a 
Newton-Raphson like method. Convergence is usually reached within 3-6 steps. The evaluation of the 
partition function \cite{M2A} in the external fields poses rather high memory demands (several Gbytes). 
Therefore we employ a CRAY T3E, assigning a subset of conformations to each processing element. Typically we 
use 64 or 128 processors in parallel, and the program scales very well with the number of processors employed\cite{M2A}. 
One needs about 1200 seconds for each set of parameters, which is roughly  2 orders of magnitude less
than for the detailed Monte Carlo simulations.

\section{ Comparison between Monte Carlo simulations and self-consistent field (SCF) calculations }
In the following we compare our Monte Carlo simulations to the results of the SCF calculations.
Both, large length scale thermodynamic properties (e.g.\ interfacial tension) as well as the local interfacial
structure (e.g.\ orientation of individual bonds) are investigated. The temperature dependence
of most quantities for symmetric blends ($f=0$) has been studied previously\cite{M4} and compared
to predictions of the Gaussian and worm-like chain model\cite{SCHMID1}. The results for vanishing bending
energy compare well to our calculations.

\subsection{Interfacial tension}
\begin{figure}[htbp]
    \begin{minipage}[t]{160mm}%
       \setlength{\epsfxsize}{13cm}
       \mbox{\epsffile{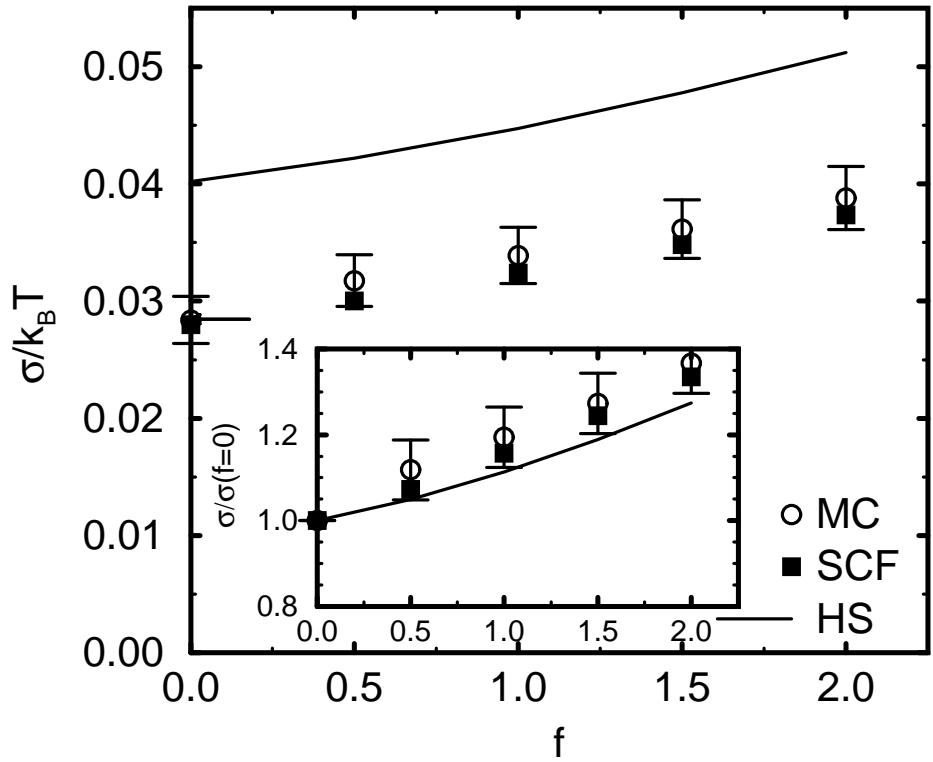}}
    \end{minipage}%
    \hfill%
    \begin{minipage}[b]{160mm}%
       \caption{
       Interfacial tensions $\sigma$ for all bending energies. Circles represent Monte Carlo estimates
       obtained from the capillary fluctuation spectrum, squares denote the values of the self consistent
       field calculations, and solid lines show the Helfand-Sapse prediction (where we have neglected the
       architecture dependence to the Flory-Huggins parameter i.e.\ $\chi=0.265$ for all $f$).
               }
       \label{fig:Ssigma}
    \end{minipage}%
\end{figure}

The interfacial tension $\sigma$ between the coexisting phases has an important impact on the morphology 
of the compound material\cite{MORPHOLOGY}. The control of domain size and shape is a key to tailoring the
application properties of the blend. The size of minority droplets often is the smaller, the smaller 
the interfacial tension between the coexisting phases\cite{MORPHOLOGY,MILNER}.
In the strong segregation limit, Helfand and Sapse\cite{HS} obtained for infinite long Gaussian chains in an 
incompressible blend the analytic expression:
\begin{equation}
\sigma = \Phi_0 \sqrt{ \chi /6 } \left( \frac{2}{3} \frac{b_A^2+b_A b_B+b_B^2}{b_A+b_B} \right)
\end{equation}
The interfacial tension $\sigma$ grows upon increasing the bending energy (i.e.\ the statistical segment length) of the 
semi-flexible component. This behavior is presented in Fig.\ \ref{fig:Ssigma}, as well as our simulation results 
and the SCF calculations, which take account of the detailed chain architecture. All data exhibit an increase of the
interfacial tension of about $30\%$.  The simulation data and the SCF calculations agree nicely on the
growth of the interfacial tension upon increasing the bending energy. The almost quantitative agreement
indicates that our identification of the $\chi$-parameter yields reasonable results for structural
asymmetric mixtures.

However, the Gaussian chain result is about a factor $1.3$ higher than the simulation data.
Recently, Ermoshkin and Semenov\cite{ER} calculated corrections to the interfacial tension due to effects of 
finite chain length $N$. For symmetric blends, they found that chain end effects reduce the interfacial tension 
by a factor $(1-4\ln2/\chi N) \approx 0.67$, which accounts well for the discrepancies between 
the Helfand-Sapse result and the Monte Carlo data. Similar reductions are found in numerical SCF 
calculations\cite{SCHMID1,SHULL}.

Note that purely entropic, packing contributions to the Flory-Huggins parameter $\Delta\chi$ are less than
$1\%$ of the total $\chi$ value for $f=1$. That is somewhat smaller than the uncertainties in identifying the
enthalpic contributions of $\chi$ and the accuracy of our interfacial tension measurement in the simulations.
Therefore, purely entropic effects derived from packing are irrelevant to the interfacial behavior for the chain lengths,
stiffness asymmetries, and temperatures investigated in the present study.

\subsection{Monomer density profiles}
\begin{figure}[htbp]
    \begin{minipage}[t]{160mm}%
       \setlength{\epsfxsize}{13cm}
       \mbox{\epsffile{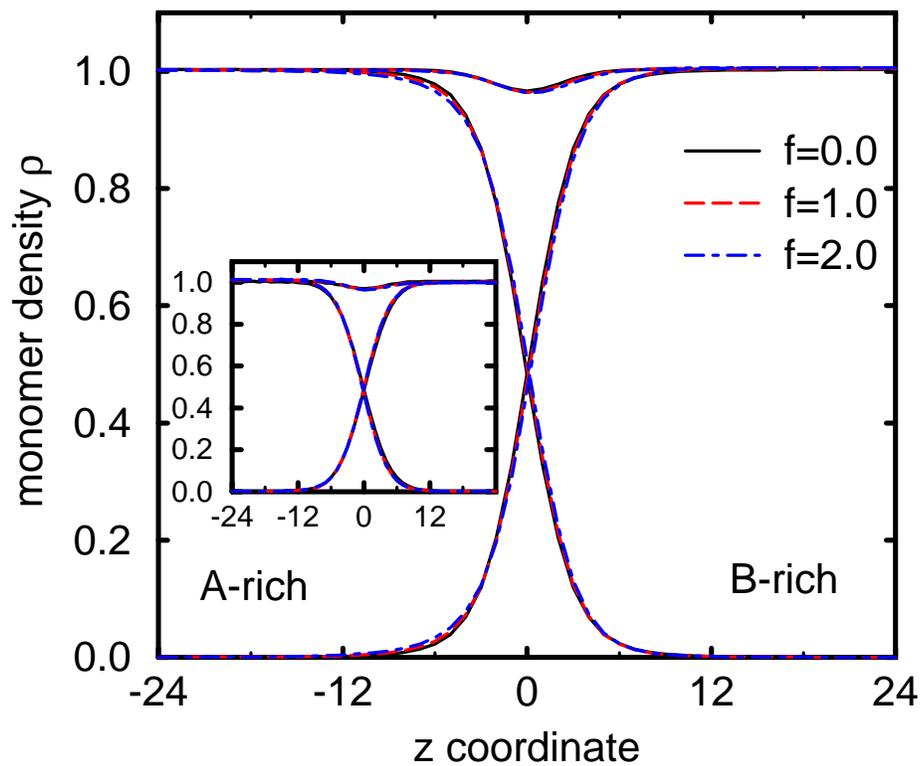}}
    \end{minipage}%
    \hfill%
    \begin{minipage}[b]{160mm}%
       \caption{
       A- and B-monomer density, and total density profiles for bending energies $f=0,1,$ and $2$
       from SCF calculations for chain length $N=32$. The apparent profiles in the Monte Carlo simulations
       are shown in the inset).
       The Monte Carlo profiles are broader than the self consistent field results, but in both cases
       there is almost no dependence on the bending energy $f$.
               }
       \label{fig:profile}
    \end{minipage}%
\end{figure}

Another important characterization of the interface are the density profiles of the individual components.
Experiments\cite{ENTANGLE} indicate, that entanglements in the interfacial zone are of major importance
for the mechanical properties of the blend. Of course, our chain lengths are too small to observe
entanglements, however static properties can be extracted from our simulation data.
The density profiles obtained from the SCF calculation are presented in Fig.\ \ref{fig:profile}, as well as the
``apparent'' profiles in the Monte Carlo simulation. The width of the apparent profile in the Monte Carlo simulations 
is about a factor 1.5 larger than the SCF result; this is not unexpected, because capillary fluctuations increase 
the squared width by a term proportional to $\ln(L)/\sigma$. However, the profiles are qualitative similar: 
both data show a 
reduction of the total monomer density at the center of the interface (the relative reduction is roughly
$\chi/2\zeta$\cite{M4}) and almost no dependence on the bending energy $f$.

\begin{figure}[htbp]
    \begin{minipage}[t]{160mm}%
       \setlength{\epsfxsize}{13cm}
       \mbox{\epsffile{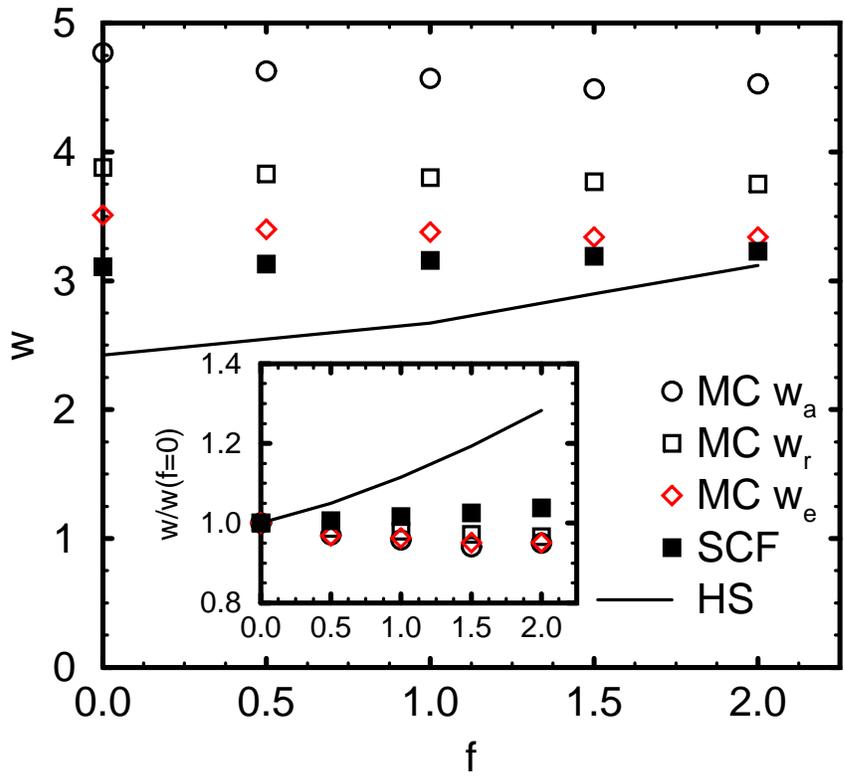}}
    \end{minipage}%
    \hfill%
    \begin{minipage}[b]{160mm}%
       \caption{
       Interfacial width as a function of bending energy $f$ for chain length $N=32$. Circles denote the apparent 
       width $w_a$ in the
       Monte Carlo simulations ($L=64$), squares represent the Monte Carlo result $w_r$ from the reduced profiles ($B=16$),
       diamonds show the estimate from the interfacial excess
       energy, filled squares mark the results of the self consistent field calculations, and the solid line
       shows the Helfand-Sapse prediction ($\chi=0.265$ for all $f$).
       The inset presents the same data normalized by the interfacial width of the symmetric blend.
               }
       \label{fig:width}
    \end{minipage}%
\end{figure}

The dependence of the interfacial width on the bending energy is shown in Fig.\ \ref{fig:width}\cite{C3}.
The width $w_r$ of the reduced profile is smaller than the apparent width $w_a$, and agrees better with the SCF results.
Due to remaining capillary wave effects it is an upper bound on the intrinsic width. All profiles presented below are
obtained by the reduced averaging procedure. The excess energy density
of the interface can also be used to estimate the intrinsic width. Since the relative increase in interfacial area due to
capillary fluctuations is of the order $\sigma\ln L/L^2$, this quantity (as well as the interfacial tension)
is not strongly affected by fluctuations of the local interfacial position. A tanh-shaped profile
$\phi_A = \frac{1}{2}(1+\tanh\frac{z}{w})$
yields in the SCF framework:
\begin{equation}
\frac{e_s}{k_BT} = \Phi_0  \int dz\; \left\{ -\frac{\epsilon z_{AA}}{2}\phi_A^2
				             -\frac{\epsilon z_{BB}}{2}\phi_B^2
					     +      \epsilon z_{AB}\phi_A\phi_B
				             \right\}
                       -\frac{\Phi_0 L (z_{AA}+z_{BB})}{4}
               \approx \frac{1}{2} w_e\Phi_0\chi
\end{equation}
where we have assumed incompressibility and neglected the finite range of interactions and any contribution of the 
intramolecular interactions to the excess energy density. Of course, this measure relies crucially on the
identification of the Flory-Huggins parameter. However, the method is computational very convenient and can be combined\cite{M3}
with the reweighting methods of measuring interfacial tensions.
It results in values which are between the reduced width and the SCF results, which shows again the consistent
parameterization of the local fluid structure. The Gaussian chain model
for $N \to \infty$ predicts for symmetric blends ($f=0$) a width which is 
about $20\%$ smaller than the SCF result.
SCF calculations\cite{SCHMID1} of Gaussian chains with the same long distance behavior
and which include chain end effects and the finite compressibility
agree within $2\%$ with our results for symmetric, flexible mixtures. An increase of the chain length from $N=32$ to $64$
reduces the effective $\chi$-parameter by $4\%$ and reduces the broadening due to finite chain length effects.
The latter effect is stronger, such that the width decreases slightly.

\begin{figure}[htbp]
    \begin{minipage}[t]{160mm}%
       \setlength{\epsfxsize}{13cm}
       \mbox{\epsffile{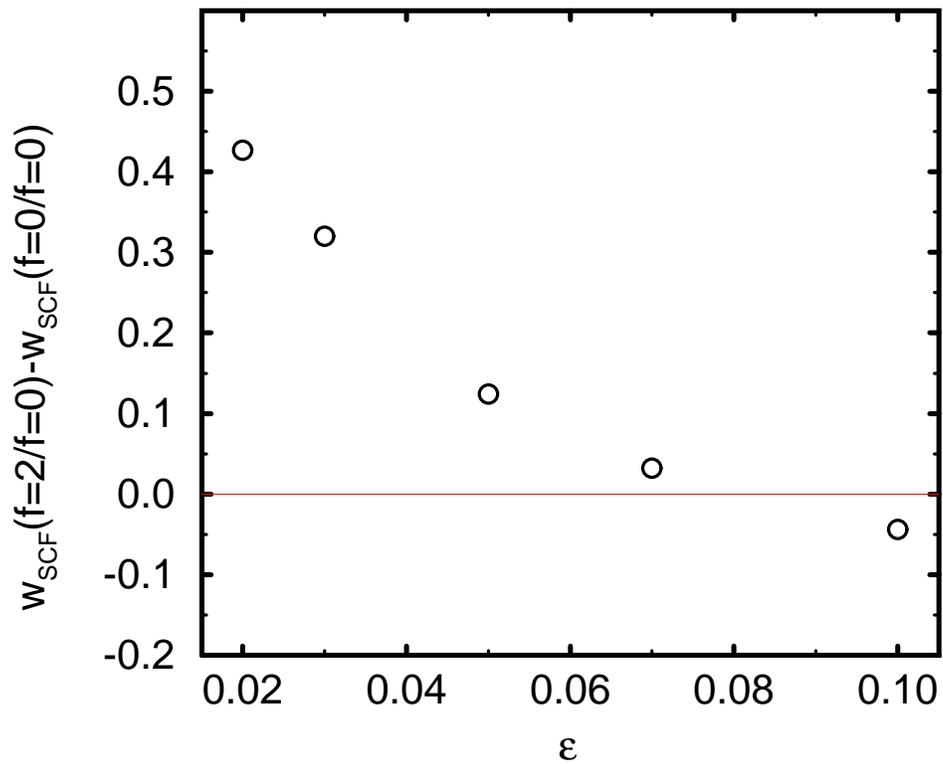}}
    \end{minipage}%
    \hfill%
    \begin{minipage}[b]{160mm}%
       \caption{
       Difference between the width of the $f=0/f=2$ blend and the symmetric $f=0$ mixture
       in the self consistent field calculations. For $\epsilon<0.082$ the interfacial width
       increases upon increasing the statistical segment length of the semi-flexible component,
       whereas at higher incompatibilities the width decreases.
               }
       \label{fig:dw}
    \end{minipage}%
\end{figure}

Most notably, the apparent width of the Monte Carlo data, the energetic width $w_e$ and the results of the SCF 
calculations show almost no dependence on the bending energy $f$, whereas the analytic expression obtained by 
Helfand and Sapse
\begin{equation}
w = \sqrt{ \frac{b_A^2+b_B^2}{12 \chi} } 
\end{equation}
predicts an increase of about $28\%$ due to the variation of $b_B$. Taking account of the stiffness dependence of
the effective Flory-Huggins parameter $\chi$, the formula above predicts an increase of $21\%$.
Qualitatively, calculations for worm-like chains\cite{MORSE,SCHMID1} indicate that increasing the bending rigidity 
results in a reduction of the interfacial width compared to the Gaussian chain result. For the present combination
of parameters, both effects seem to cancel, resulting in an interfacial width, which is nearly independent of the 
bending rigidity. For lower incompatibilities, the interfacial width is larger, the Gaussian description on the length
scale of the interfacial width becomes more appropriate. Therefore the difference between the width in the
flexible/semi-flexible blend and the width in the symmetric flexible mixture increases
in accord with the Helfand Sapse description. This is confirmed by SCF calculations (c.f.\ Fig.\ \ref{fig:dw}), where 
we have assumed that the effective coordination numbers are temperature independent. However, upon increasing the incompatibility
further ($\epsilon>0.082$), one finds that an increase of the statistical segment length results in a {\em smaller} interfacial width
of the asymmetric blend in qualitative contrast to the predictions of the Gaussian model.

\subsection{Distribution of chain ends and orientations}
\begin{figure}[htbp]
    \begin{minipage}[t]{160mm}%
       \setlength{\epsfxsize}{9cm}
       \mbox{\epsffile{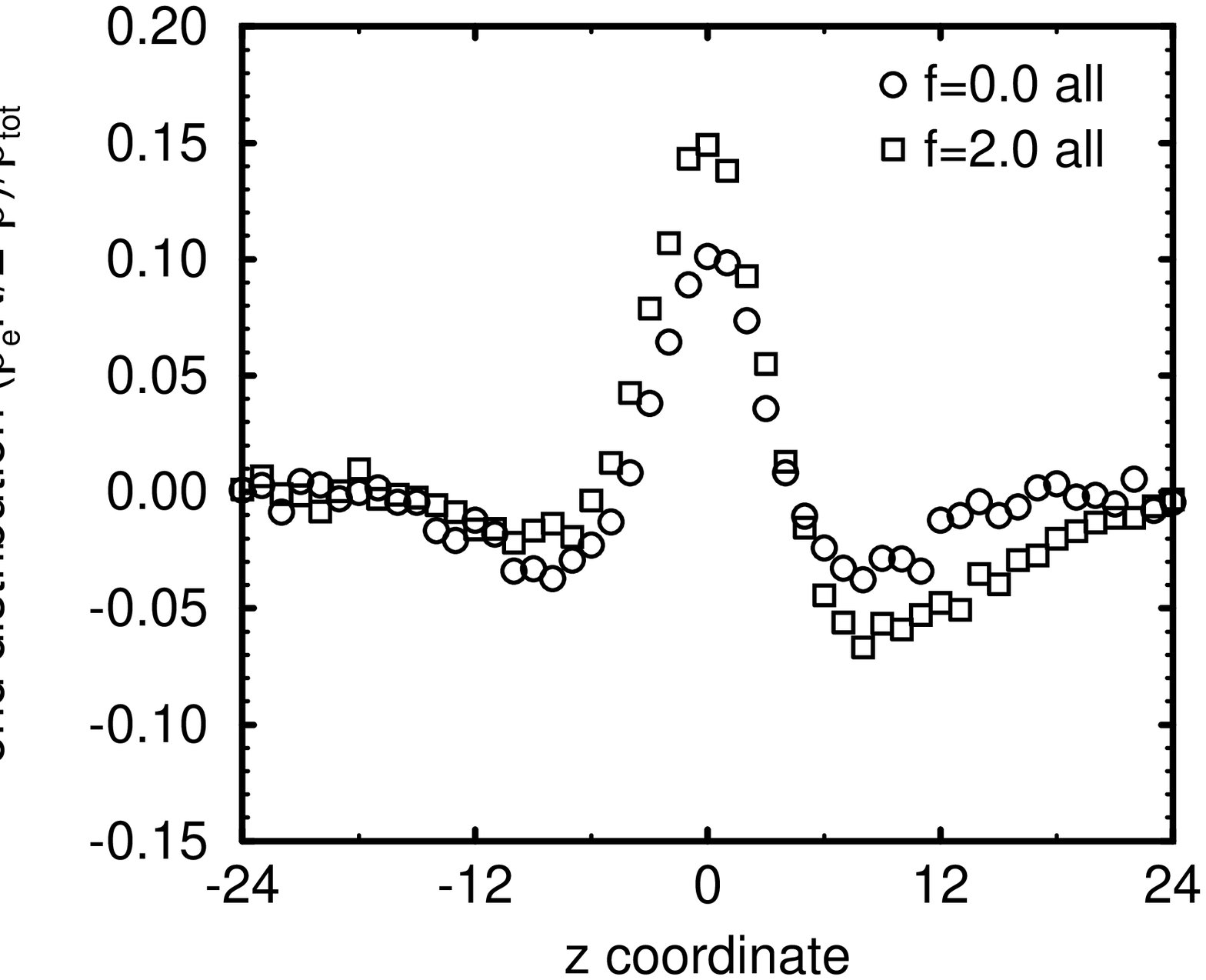}}
       \setlength{\epsfxsize}{9cm}
       \mbox{\epsffile{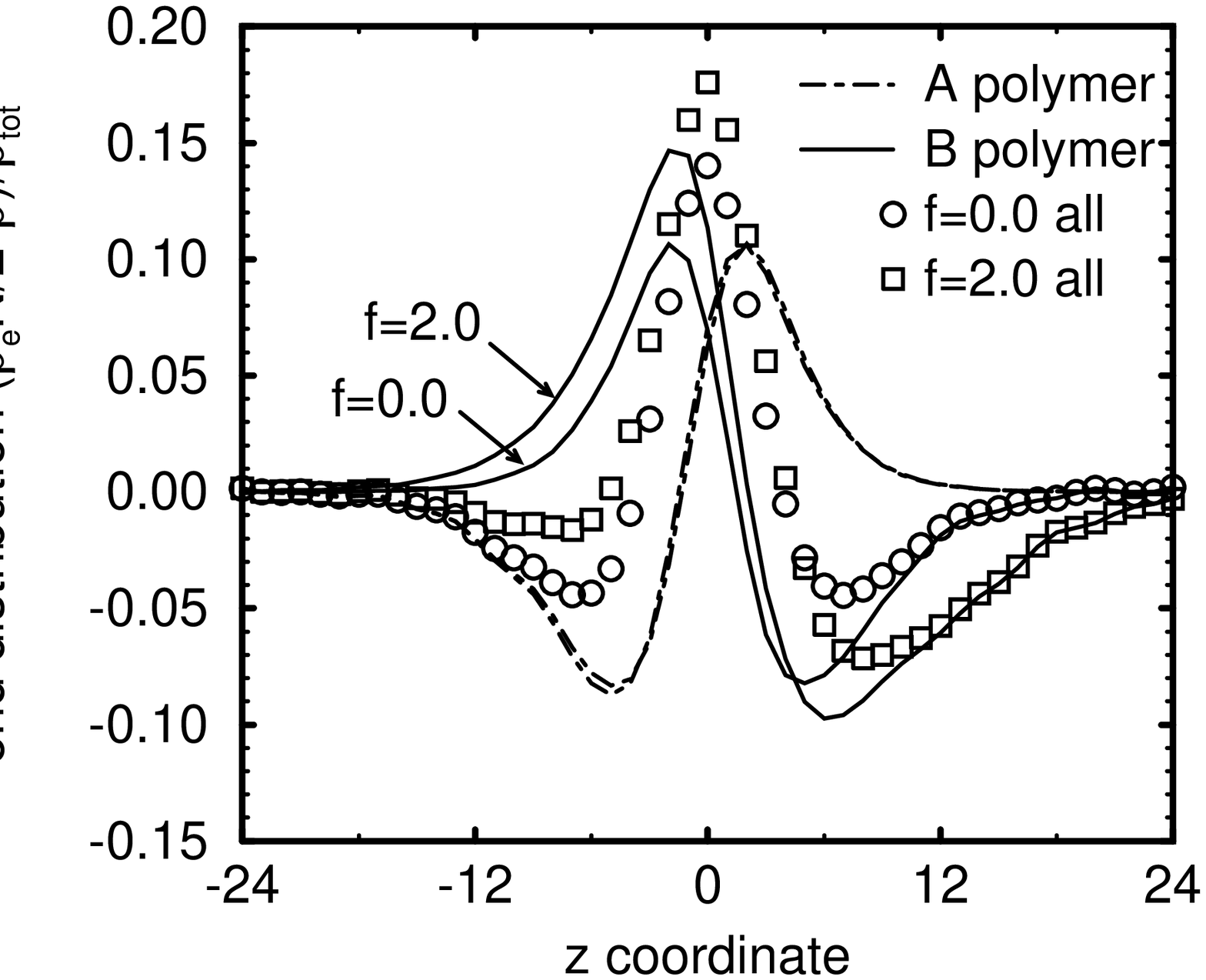}}
       \setlength{\epsfxsize}{9cm}
       \mbox{\epsffile{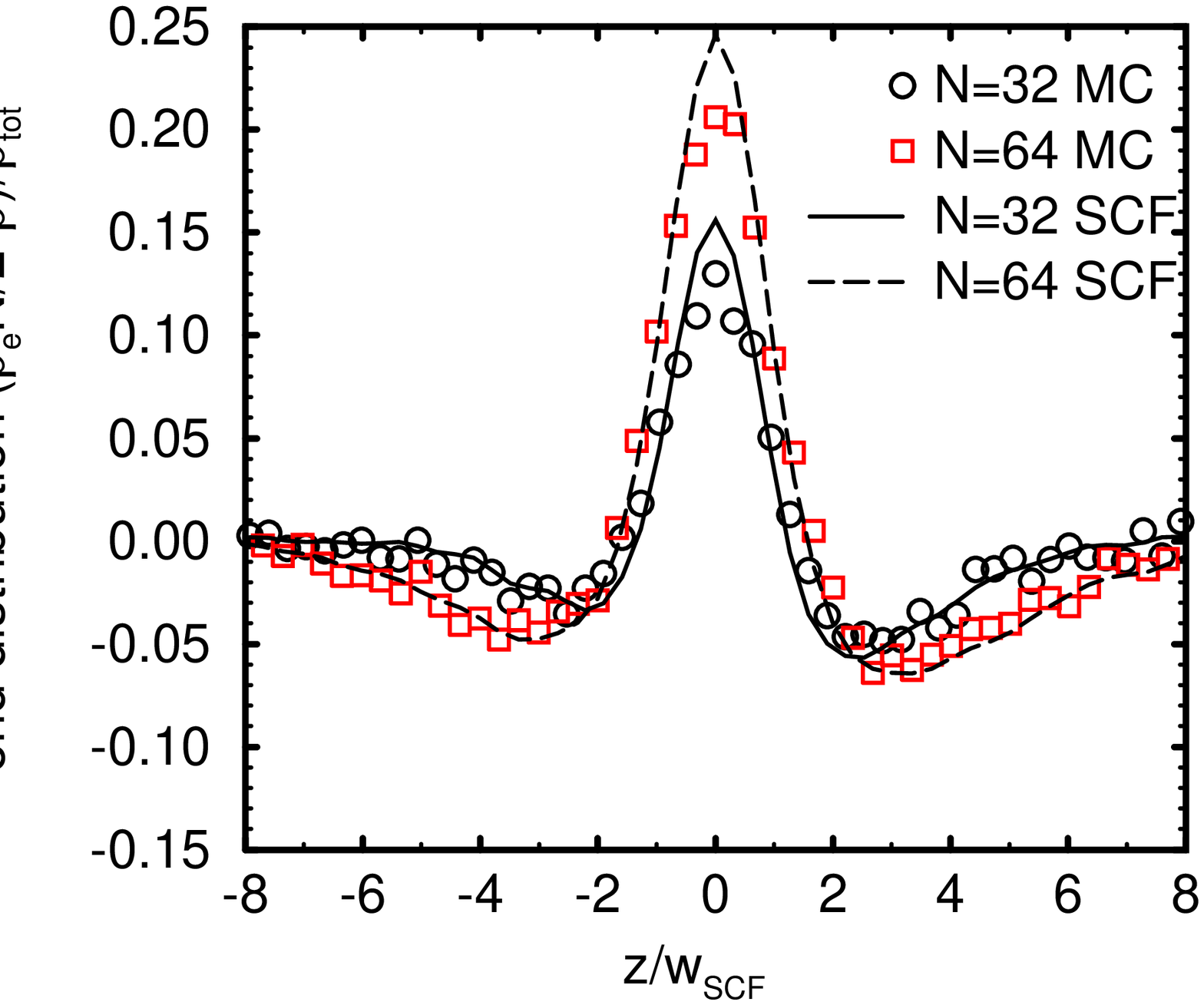}}
    \end{minipage}%
    \hfill%
    \begin{minipage}[b]{160mm}%
       \caption{
       Enrichment of chain ends $\frac{\rho_eN/2-\rho}{\rho_{\rm tot}}$
       ($\rho_e$: density of chain ends, $\rho$ the corresponding monomer density, and 
       $\rho_{\rm tot}$ total monomer density)
       at the interface for bending energies $f=0$ and $2$.
       ({\bf a}) Monte Carlo results and ({\bf b}) self consistent field calculations
       ({\bf c}) chain lengths $N=32$ and $64$ for $f=1$.
               }
       \label{fig:end}
    \end{minipage}%
\end{figure}

The enrichment of chain ends at the center of the interface\cite{M4} and at hard walls\cite{SKKUMAR,BITSANIS,SCHMID,JOERG} has attracted abiding 
interest. Chain ends are important 
for the interdiffusion and healing properties at interfaces between long polymers\cite{WU}. They also play an important 
role for reactions at interfaces. In many experimental systems, chain ends have slightly different interactions
than inner chain segments,which might result in a modification of the interfacial properties. On the theoretical
side, the behavior of chain ends is related to corrections to the ground state approximation. Therefore it is a
sensitive test for a quantitative theoretical description. Chain end effects give rise
to large corrections to the interfacial width and tension, and they also play an important role for long range interactions
between interfaces\cite{ER}. The distribution of chain ends for symmetric blends has been investigated by
Monte Carlo simulations\cite{M4}, and in the framework of SCFT for Gaussian chains\cite{WU,SCHMID1}.
In Fig.\ \ref{fig:end} the simulational results and the SCF calculations are presented; both agree almost
quantitatively. As in symmetric blends, chain ends are enriched at the center of the interface, and this 
effect goes along with a depletion away from the interface. The fact that the depletion zone in the wings shifts outwards
with increasing chain length, indicates that the length scale of the rearrangement of chain ends is the radius of gyration.
A-polymers stick their ends into the B-rich phase and vice versa. The effect on the semi-flexible chains becomes 
more pronounced with growing stiffness, while the A-polymers are hardly influenced by the stiffness of the B-polymers.

\begin{figure}[htbp]
    \begin{minipage}[t]{160mm}%
       \setlength{\epsfxsize}{9cm}
       \mbox{\epsffile{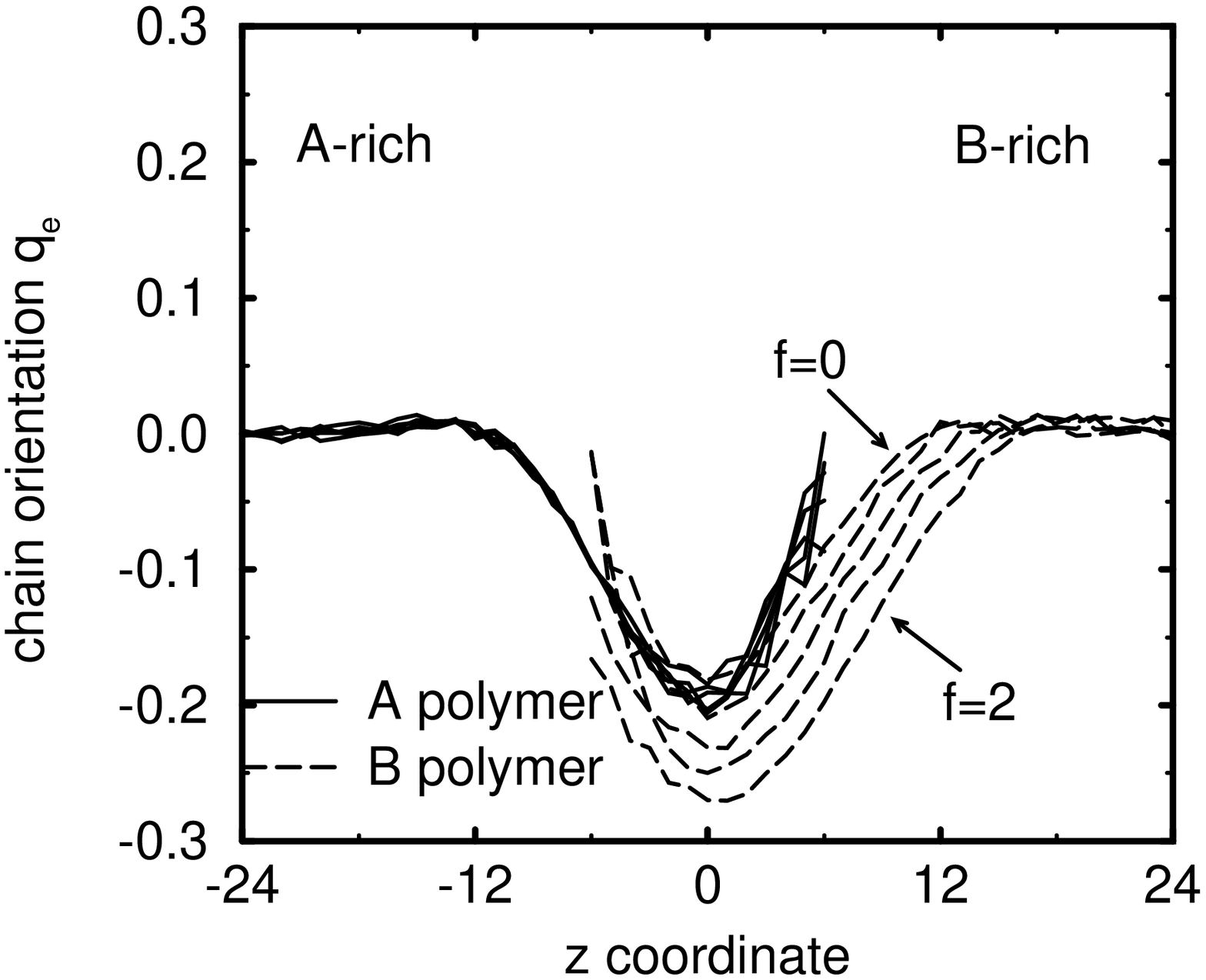}}
       \setlength{\epsfxsize}{9cm}
       \mbox{\epsffile{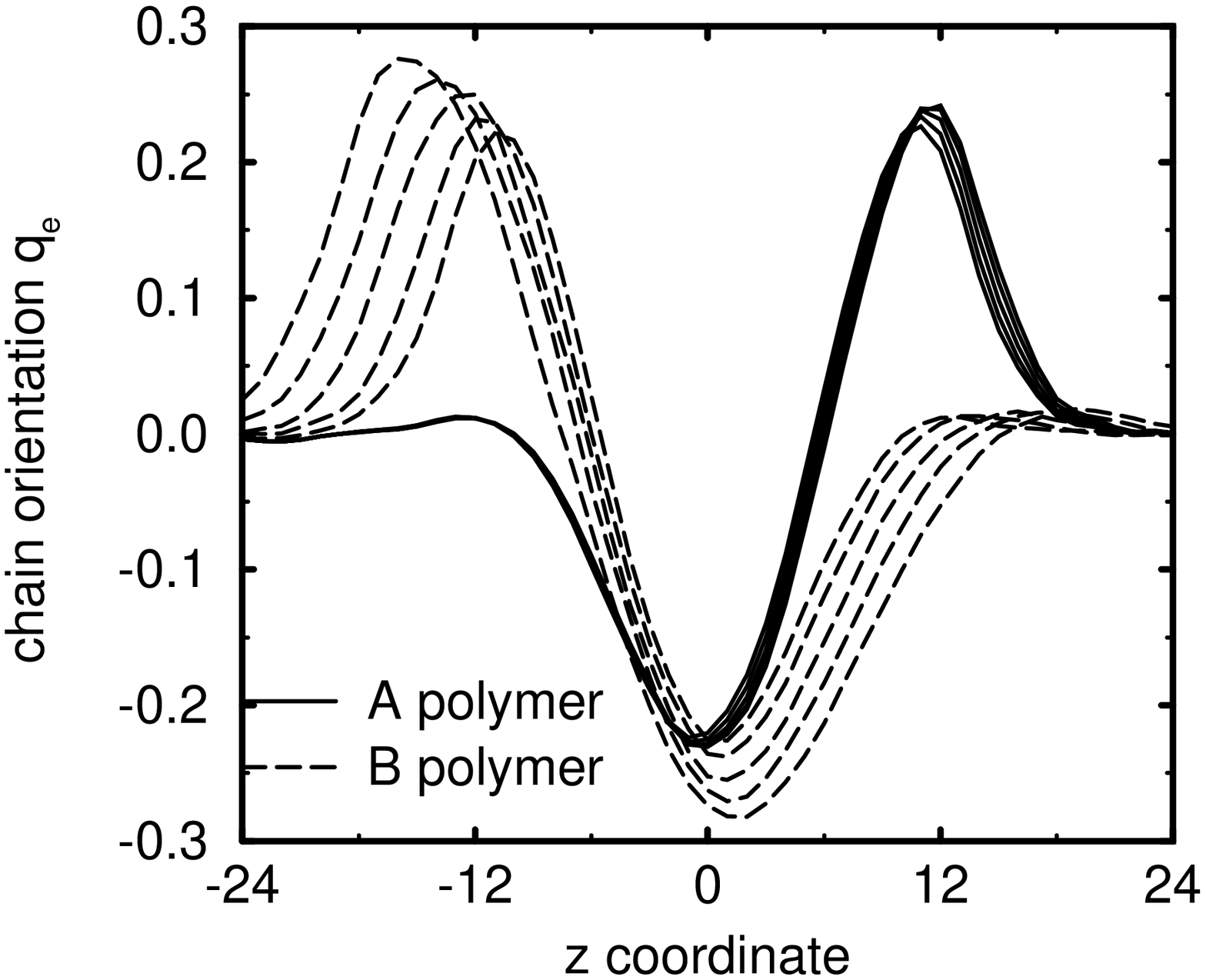}}
       \setlength{\epsfxsize}{9cm}
       \mbox{\epsffile{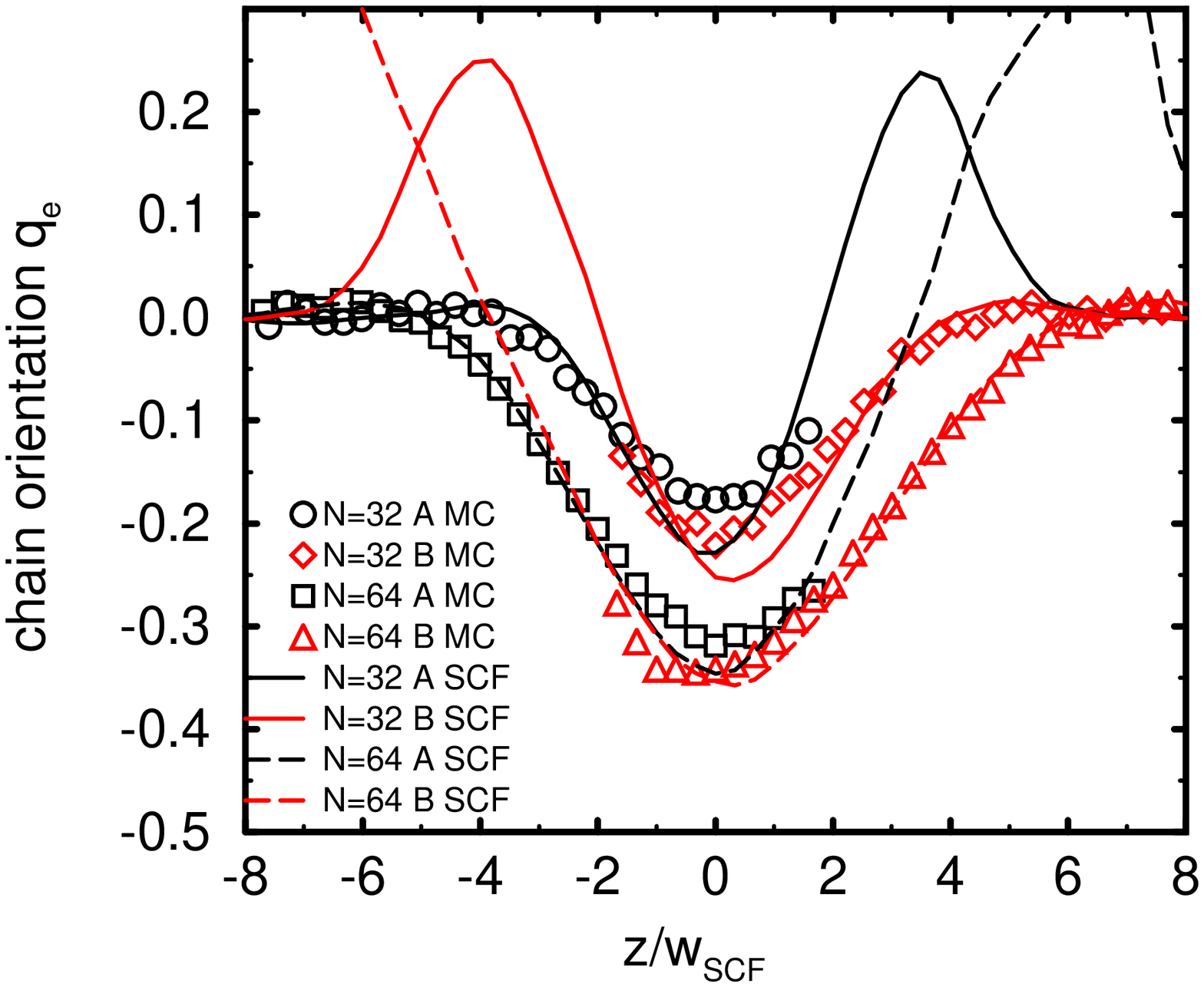}}
    \end{minipage}%
    \hfill%
    \begin{minipage}[b]{160mm}%
       \caption{
       Orientation of the end-to-end vector $q_e$, for A-polymers (solid lines) and B-polymers
       (dashed lines) and all bending energies $f$. The orientation of the B-polymers increases
       upon increasing $f$.
       ({\bf a}) Monte Carlo results and ({\bf b}) self consistent field calculations. 
       ({\bf c}) chain lengths $N=32$ and $64$ for $f=1$.
               }
       \label{fig:qe}
    \end{minipage}%
\end{figure}

The instantaneous shape of a polymer coil is a prolate ellipsoid\cite{M4}.
Polymers orient themselves by putting their ends preferentially at the center of the interface. This is quantified
by the orientational parameter\cite{M4} for the end-to-end vector (cf.\ Fig.\ \ref{fig:qe}):
\begin{equation}
q_e(z) = \frac{3\langle R^2_z\rangle_z-\langle \vec{R}^2\rangle_z}{2\langle \vec{R}^2\rangle_z}
\end{equation}
where the outer index $z$ at the brackets denotes the z coordinate of the midpoint of the end-to-end vector $\vec{R}$,
and the inner indices its Cartesian components. The chains align their two long axis parallel to the interface in
their majority phases, similar to the behavior at a hard wall. The chain orientation of semi-flexible polymers 
increases for growing stiffness, while the flexible A-polymers are not affected. The agreement between Monte Carlo simulation 
and SCF calculations is again almost quantitative. In the SCF framework, the orientations of the chains in the 
minority phase is accessible. The polymers align perpendicular to the interface, as to reach with one end their 
corresponding bulk phase. The length scale of the ordering increases with the bending energy $f$ and with chain length $N$.
\newline
The orientation of individual bond vectors $q_b$ shows a similar behavior. Bonds align parallel to the interface;
the effect for the semi-flexible component grows with increasing bending energy and its range is largely independent
of the chain length. The agreement between simulations and SCF calculations is very good. The Gaussian chain model 
cannot predict any nonzero orientation of the bonds. The orientation of bonds in our model is, in fact, much smaller than for the 
end-to-end distance\cite{M4}. 

In contrast to the width of the density profile, the spatial range over which the orientation of bonds 
extends grows upon increasing the bending energy. Therefore, the orientational width and the width of
the composition profile are  two independent microscopic length scales.

\section{ Conclusions and outlook} 
In summary, we have presented extensive simulations of highly incompatible polymers with 
different stiffness. The local structure of the interface has been characterized by density profiles
of different monomer species and chain ends and orientational profiles of whole chains and individual 
bonds. The interfacial tension has been measured via analyzing the spectrum of capillary fluctuations.
Using the pair correlation functions of the pure components and a random-packing
like assumption for the intermolecular contacts between different species, we have extracted
an effective Flory-Huggins parameter, which takes account of the stiffness dependence of the structure
of the polymeric fluid. The effective Flory-Huggins parameter grows upon increasing the stiffness, because
 back folding is less probable and the number of intermolecular contacts increases respectively.

This effective Flory Huggins parameter was then employed in SCF calculations, as well as the
chain conformations in the pure melt. These calculation incorporate the chain structure on {\em all} length scales via
a partial enumeration scheme; there is no free parameter in describing the chain architecture. 
Using the detailed local structure of the bulk (as obtained by simulations) in the SCF calculations, we predict the interfacial
properties.

Monte Carlo results and SCF calculations for the interfacial tension, the excess interfacial energy, 
the redistribution of chain ends and orientations of whole chains and individual bonds agree very well provided that the analysis accounts for capillary fluctuations.
However, comparing our results to the analytical predictions of the Gaussian chain model for infinite chain length, 
we find qualitative deviations, especially for the dependence of the interfacial width on the chain stiffness. 
This finding might be important for extracting the Flory-Huggins parameter from interfacial profiles in highly
incompatible polymer blends. Therefore, our results emphasize that the local structure, both of the underlying 
monomer fluid and of the chain architecture, is important for a quantitative description.

The radius of gyration determines the range of orientation of whole chains and the distribution of chain ends.
Furthermore, we identify  two independent microscopic length scales of the interfacial profile; one controls the 
width of the monomer density profile, the other corresponds to the range of orientations.
This behavior resembles the findings in symmetric blends of worm-like chains in the limit 
$\kappa\chi \gg 1$\cite{MORSE} and the behavior of a homopolymer melt at a hard wall
which is the limiting case for infinite incompatibility. However, in the present study this behavior is found 
for a different model
which can be described neither by Gaussian nor by worm-like statistics on small length scales.
Deviations from the Gaussian model occur under rather mild conditions which correspond roughly to 
$\kappa\chi=0.18 \cdots 0.47$ in the equivalent worm-like chain model.
Furthermore our self consistent field approach as well as the simulation techniques are applicable to
arbitrary chain architecture\cite{M2}.

Assuming that the chain conformations and the local fluid structure
are approximately independent of temperature, we have extended the self consistent field calculations
to other incompatibilities. The results indicate that chain architecture becomes important when its length
scale is comparable with the interfacial width. At very high incompatibility, increasing the stiffness of the
semi-flexible component results in a decrease of the interfacial width.  However, the Gaussian chain results 
and our calculations, which take account of the explicit chain architecture on all length scales, agree better for lower 
incompatibilities, where the interfacial width is much larger than the persistence length.

\subsection*{Acknowledgment}
It is a great pleasure to thank K. Binder, G.S. Grest and F. Schmid for helpful and stimulating discussion,
and M. Schick for critical reading of the manuscript. Generous access to the CRAY T3E at the San Diego 
Supercomputer Center (through a grant to M. Schick) is also gratefully acknowledged. M.M. thanks the Bundesministerium
f\"ur Forschung, Technologie, Bildung und Wissenschaft(BMBF) for support under grant No. 03N8008C.
A.W. thanks the Deutsche
Forschungsgemeinschaft for support under grant number Bi 314/3.

\pagestyle{empty}

\begin{figure}[htbp]
    \begin{minipage}[t]{160mm}%
       \setlength{\epsfxsize}{9cm}
       \mbox{\epsffile{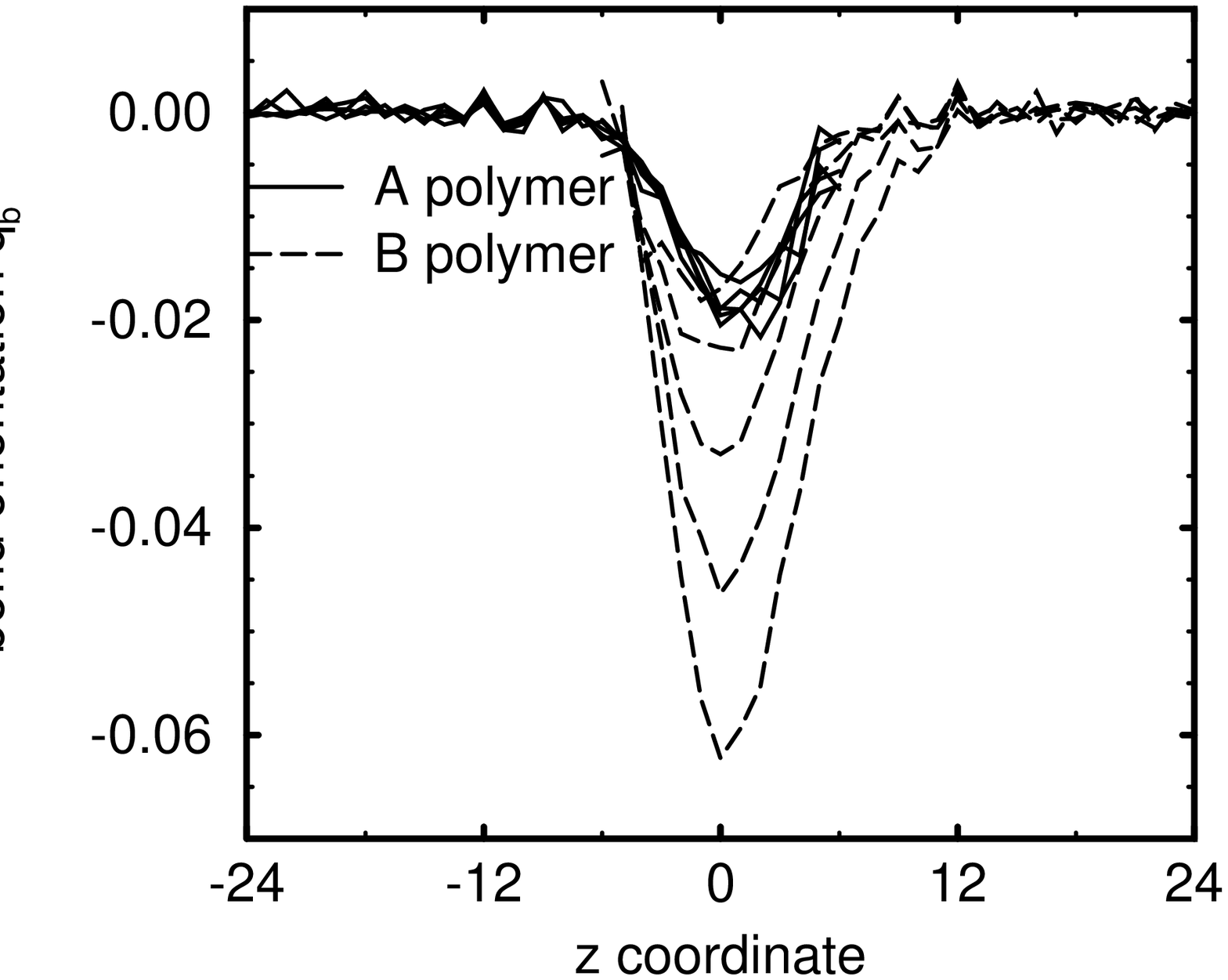}}
       \setlength{\epsfxsize}{9cm}
       \mbox{\epsffile{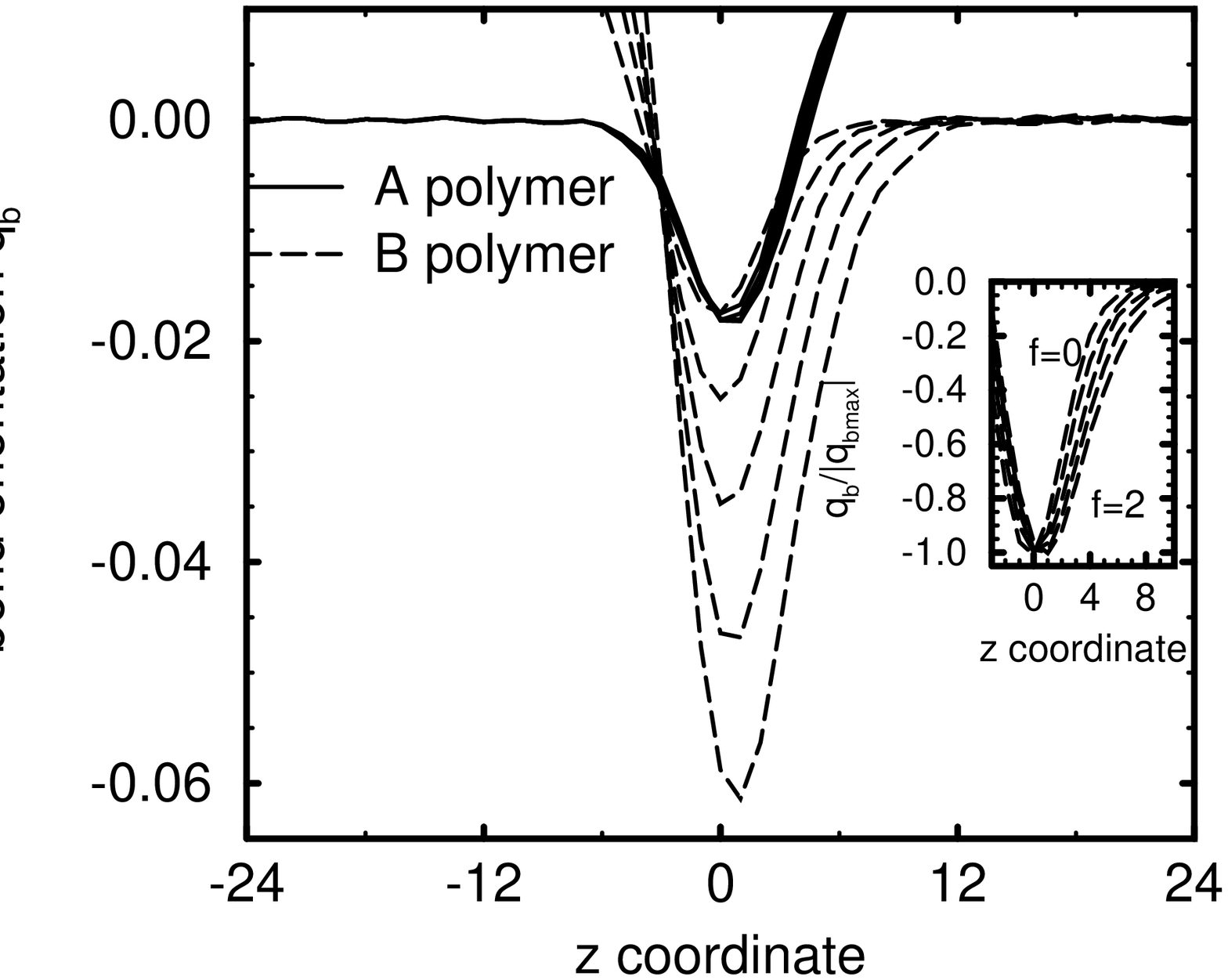}}
       \setlength{\epsfxsize}{9cm}
       \mbox{\epsffile{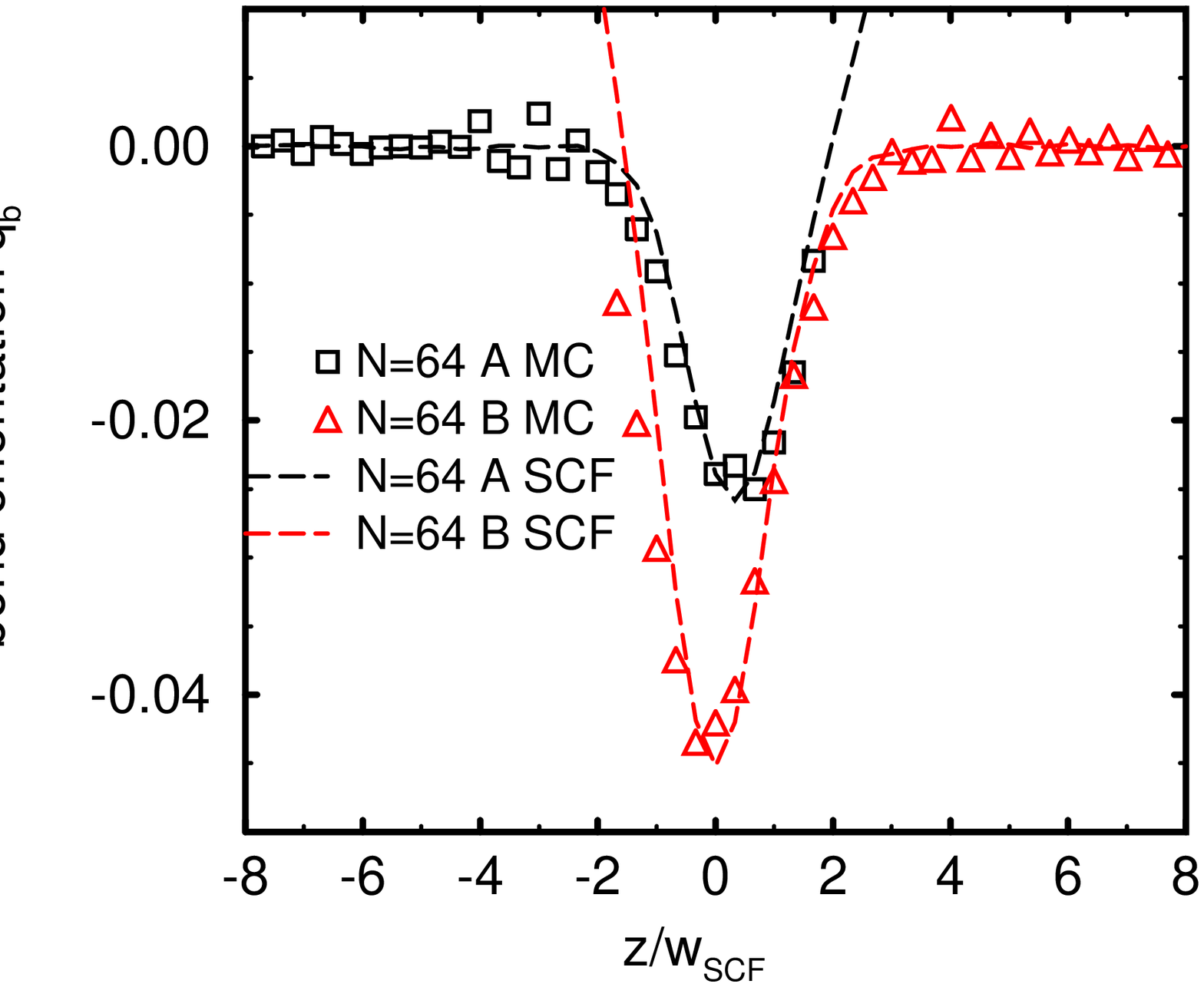}}
    \end{minipage}%
    \hfill%
    \begin{minipage}[b]{160mm}%
       \caption{
       Orientation of bond vectors $q_b$, for A-polymers (solid lines) and B-polymers
        $q_b(z) = \frac{3\langle b^2_z\rangle_z-\langle \vec{b}^2\rangle_z}{2\langle \vec{b}^2\rangle_z}$
       for A-polymers (solid lines) and B-polymers
       (dashed lines) and all bending energies $f$. The orientation of the B-bonds and the length scale of ordering
       increases upon increasing $f$.
       ({\bf a}) Monte Carlo results and ({\bf b}) self consistent field calculations
       The inset presents
       the self-consistent field results normalized to the maximum of $q_b$. Note, that the length scale
       of the orientation grows upon increasing $f$.
       ({\bf c}) chain length $N=64$ and $f=1$.
               }
       \label{fig:Sqb}
    \end{minipage}%
\end{figure}

\begin{table}[htbp]
\begin{tabular}{|c|c|c|c|c|c|c|c|c|c|c|c|} 
\hline
$N$   &   $f$   &   $\langle b^2\rangle$  &  $R^2$  & $R_g^2$ & $z_{BB}$  &  $\langle e \rangle/k_BT$  &  $e_s/k_BT$  & $w_a$ & $w_r$ & $w_e$  &  $w_{\rm scf}$\\
\hline
\hline
32    &   0.0   &   6.92  &  290.4   &  48.8  &  2.65  &  -0.00732  &  0.0290  &  4.77  &  3.88  &  3.51  &  3.11  \\
      &   0.5   &   6.88  &  350.1   &  58.1  &  2.84  &  -0.00731  &  0.0293  &  4.63  &  3.83  &  3.40  &  3.13  \\
      &   1.0   &   6.86  &  431.8   &  70.8  &  3.02  &  -0.00730  &  0.0300  &  4.57  &  3.80  &  3.38  &  3.16  \\
      &   1.5   &   6.84  &  536.9   &  86.5  &  3.17  &  -0.00730  &  0.0304  &  4.49  &  3.77  &  3.34  &  3.19  \\
      &   2.0   &   6.84  &  665.2   &  105.1 &  3.29  &  -0.00730  &  0.0310  &  4.53  &  3.75  &  3.34  &  3.23  \\
\hline
64    &   0.0   &   6.92  &  609.3   &  101.7 &  2.53  &  -0.00732  &          &        &        &        &        \\
      &   1.0   &   6.86  &  987.3   &  148.3 &  2.92  &  -0.00730  &  0.0265  &  4.13  &  3.43  &  3.11  &  2.99  \\ 
\end{tabular}    
\caption{Single chain conformations and interfacial properties as a function of the bending energy $f$.
Interfacial data refer to blends of flexible ($f=0$) and semi-flexible ($f$ as indicated) chains.
$\langle b^2\rangle$: mean squared bond length, $R^2$: mean squared end-to-end distance, 
$R_g^2$ mean squared radius of gyration, $z_{BB}$ effective intermolecular coordination number as measured in 
simulations of the bulk system,
$\langle e \rangle/k_BT$: bulk energy density, 
$e_s/k_BT$: interfacial energy excess per unit area, $w_a$ apparent interfacial with for $L=64$,
$w_r$ reduced interfacial width for block size $B=16$, $w_e$ estimated interfacial with from the excess energy,
$w_{\rm scf}$ interfacial width in the SCF calculations.}
\label{tab:konf}
\end{table}



\begin{thebibliography}{99}
\bibitem{C3}         Upon increasing the stiffness disparity, the profiles of our compressible blend
		     become slightly asymmetric. Fitting a tanh-profile to the A- and B-monomer density,
		     one finds that the B-profile is about $1\%$ wider ($f=2$) than the A-profile.
		     This effect is observed in simulations and SCF calculations.
\end{thebibliography}
\end{document}